\documentclass[twocolumn,showpacs,preprintnumbers,amsmath,amssymb,prb]{revtex4}

\usepackage{graphicx}
\usepackage{dcolumn}
\usepackage{bm}

\begin{document}


\title{Mapping of strongly correlated steady-state nonequilibrium 
to an effective equilibrium}

\author{J. E. Han}
\affiliation{
Department of Physics, State University of New York at Buffalo, Buffalo, NY 14260, USA}

\date{\today}

\begin{abstract}
By mapping steady-state nonequilibrium to an effective equilibrium,
we formulate nonequilibrium problems within an equilibrium
picture where we can apply existing equilibrium many-body techniques to
steady-state electron transport problems. We study the analytic
properties of many-body scattering states, reduce
the boundary condition operator in a simple form and prove that
this mapping is equivalent to the correct linear-response theory.
In an example of infinite-$U$ Anderson impurity model,
we approximately solve for
the scattering state creation operators, based on which we 
derive the bias operator $\hat{Y}$ to construct the nonequilibrium
ensemble in the form of the Boltzmann factor $e^{-\beta(\hat{H}
-\hat{Y})}$. The resulting Hamiltonian is solved by the non-crossing
approximation. We obtain the Kondo anomaly conductance at zero bias,
inelastic
transport via the charge excitation on the quantum dot and
significant inelastic current background over a wide range of bias.
Finally, we propose a self-consistent algorithm of mapping
general steady-state nonequilibrium.
\end{abstract}


\maketitle

\section{Introduction}

Quantum transport theory~\cite{datta_book,rammer,wingreen}
has recently been one of the most exciting
fields of condensed matter physics. The field not only has a great
promise for providing theoretical framework for the design 
of nanoscale electronic devices, but also has been a frontier
of quantum many-body theory in nonequilibrium.

In regards to predicting behavior of nano-electronic devices
the current theories based on non-equilibrium Green function 
technique~\cite{wingreen,datta_prb} 
have been quite successful. However, as we approach 
strongly correlated
regime of transport~\cite{wingreen,langreth} 
under extreme nonequilibrium conditions,
approximate Green function techniques become 
increasingly unreliable.
One can make an analogy to the equilibrium quantum 
statistical mechanics, where the combination of diagrammatic methods 
and computational techniques has been vital for the great success
in the quantum many-body theory during the past decades.

Therefore, it is a pressing issue that we have an alternative
theoretical scheme to implement the computational techniques to 
nonequilibrium problems. Recently new approaches have been 
proposed to solve nonequilibrium. Mehta and Andrei~\cite{mehta}
have applied the non-equilibrium Bethe Ansatz technique to
solve the interacting resonant-level model. Renormalization
group theory has been extended to nonequilibrium 
systems~\cite{rosch} to investigate the time-dependence~\cite{anders}.

One of the main problems in applying
numerical techniques to transport problems
has been the lack of understanding of
nonequilibrium ensemble, specifically, as to how nonequilibrium
boundary conditions can be implemented as a statistical operator.
In steady-state nonequilibrium, Zubarev~\cite{zubarev} has extended
the Gibbsian statistical mechanics to incorporate 
steady-state boundary conditions in the density matrix formalism.
Later, Hershfield~\cite{hershfield} has shown that the nonequilibrium
ensemble can be expressed in the effective Boltzmann factor
$e^{-\beta(\hat{H}-\hat{Y})}$
with the bias operator, $\hat{Y}$, in terms of the
scattering state operators. Once the effective
Hamiltonian $\hat{H}-\hat{Y}$
inside the Boltzmann factor is obtained, one can 
use the equilibrium numerical techniques to sample the
nonequilibrium ensemble.

Despite the great potential, application of the method has been
limited due to the difficulties in constructing the bias
operator~\cite{schiller}. Recently, the method has been implemented
by the author~\cite{elph}
in the system of quantum dot (QD) coupled with electron-phonon interaction.
In the work, he has approximately constructed the bias operator
$\hat{Y}$ by truncating the non-local interaction arising from the
nonequilibrium boundary condition and by expanding the scattering state
operators up to the harmonic order of the electron-phonon interaction.
However, in the strongly correlated limit, such perturbative expansion of
the scattering state may not be sufficient to reproduce the many-particle
transport such as the Kondo anomaly conductance.

Ultimately, it is desirable that we develop a general computational
scheme of mapping nonequilibrium system to an effective equilibrium
Hamiltonian, so that equilibrium techniques such as quantum Monte Carlo
method can be used to directly calculate nonequilibrium properties.
However, to achieve the goal, we must have a good understanding of
the properties of the nonequilibrium ensemble. Therefore, 
in order to gain physical insights of the mapping, we
first study the diagrammatic construction of the nonequilibrium
ensemble.

The first main goal of this paper is to derive general analytical properties
of scattering state operators, especially the anti-commutation
and completeness
relations in the interacting limit, and derive various expressions 
of the boundary condition operator. By clarifying the fundamental
questions on the properties of scattering state operators, we 
lay a foundation of the mapping the nonequilibrium. Furthermore,
we show
that the nonequilibrium mapping produces the correct theory
in the zero-bias limit.

The second goal is, by taking an example, to address what 
correlation effects
have to be included in the bias operator $\hat{Y}$ to properly describe
the strongly correlated transport. To this end, we consider the
Anderson impurity Hamiltonian as a model for Kondo dot systems. 
We might expect that, since the original Hamiltonian has a strong on-site
interaction, the strong correlation effects will be described simply
by shifting chemical potentials in the source, drain reservoirs.
However, it will be shown in this work that it is essential to include
the correlation effects at the level of Hamiltonian in the boundary
condition to produce the Kondo anomaly~\cite{kondo}. 

This paper is organized as follows. In Section~\ref{sec:analytic}, 
we introduce
the idea of mapping the nonequilibrium and present the analytic
properties of
the scattering state and the bias operators. The proofs
are provided in Appendices. In the following
Section~\ref{sec:anderson}-A, we 
develop a procedure to expand the scattering state operator
using the slave-boson representation of the QD states in the
infinite on-site Coulomb interaction. We derive expression
for the scattering state operators and give physical interpretations
in the expansion. In the next subsection~\ref{sec:anderson}-B, 
we construct the bias
operator $\hat{Y}$ and explain how the nonequilibrium Hamiltonian
is calculated in the non-crossing approximation in the equilibrium
picture. In the subsection~\ref{sec:anderson}-C, 
we derive an approximate
expression for the current. In section~\ref{sec:results}, 
we present numerical 
results and discuss strong correlation effects in the differential
conductance.
In section~\ref{sec:sc}, we discuss how the limitations in the
previous section can be improved in a self-consistent algorithm.

\section{Analytic Properties of the Mapping}
\label{sec:analytic}

\begin{figure}[bt]
\rotatebox{0}{\resizebox{!}{1.6in}{\includegraphics{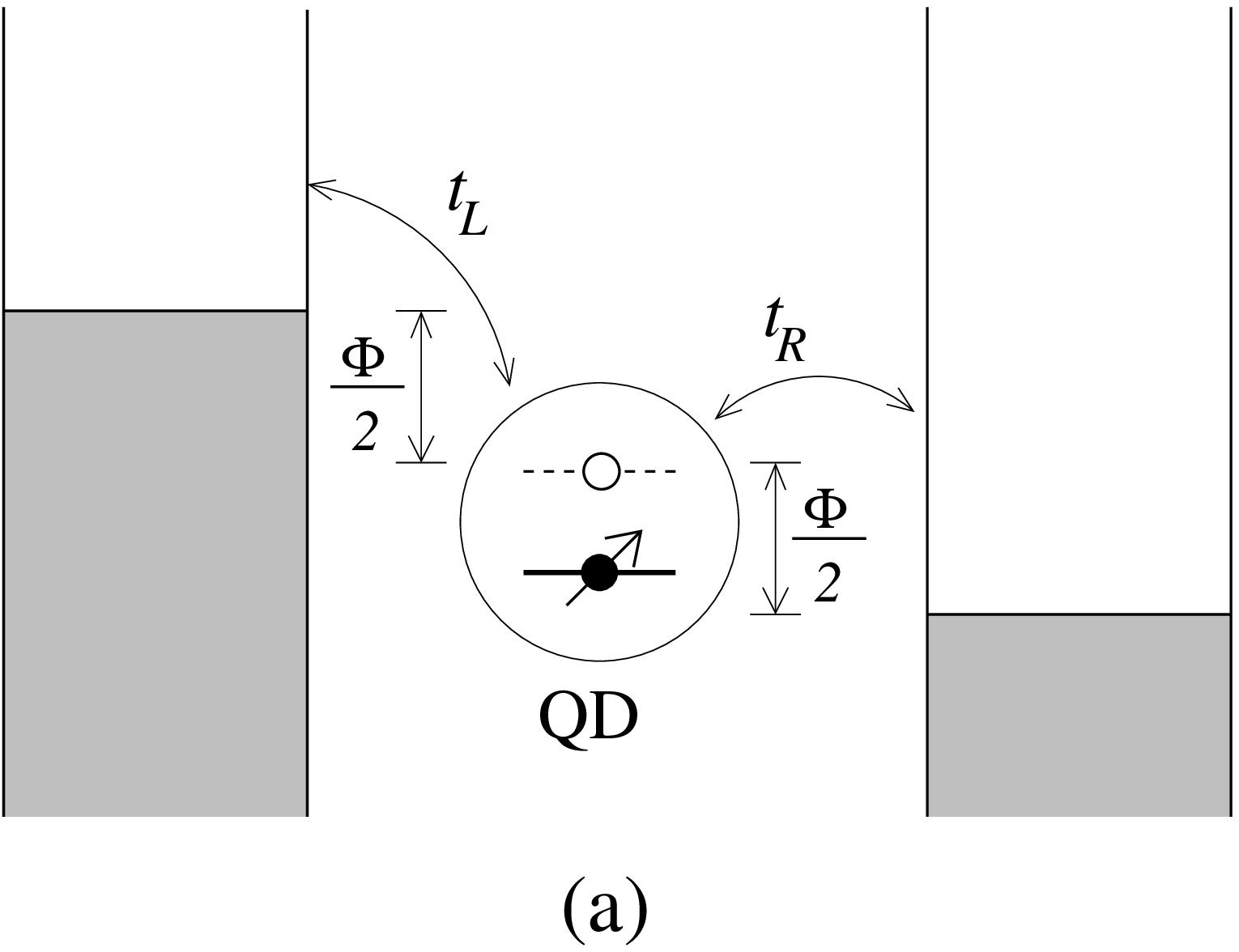}}}
\rotatebox{0}{\resizebox{!}{1.6in}{\includegraphics{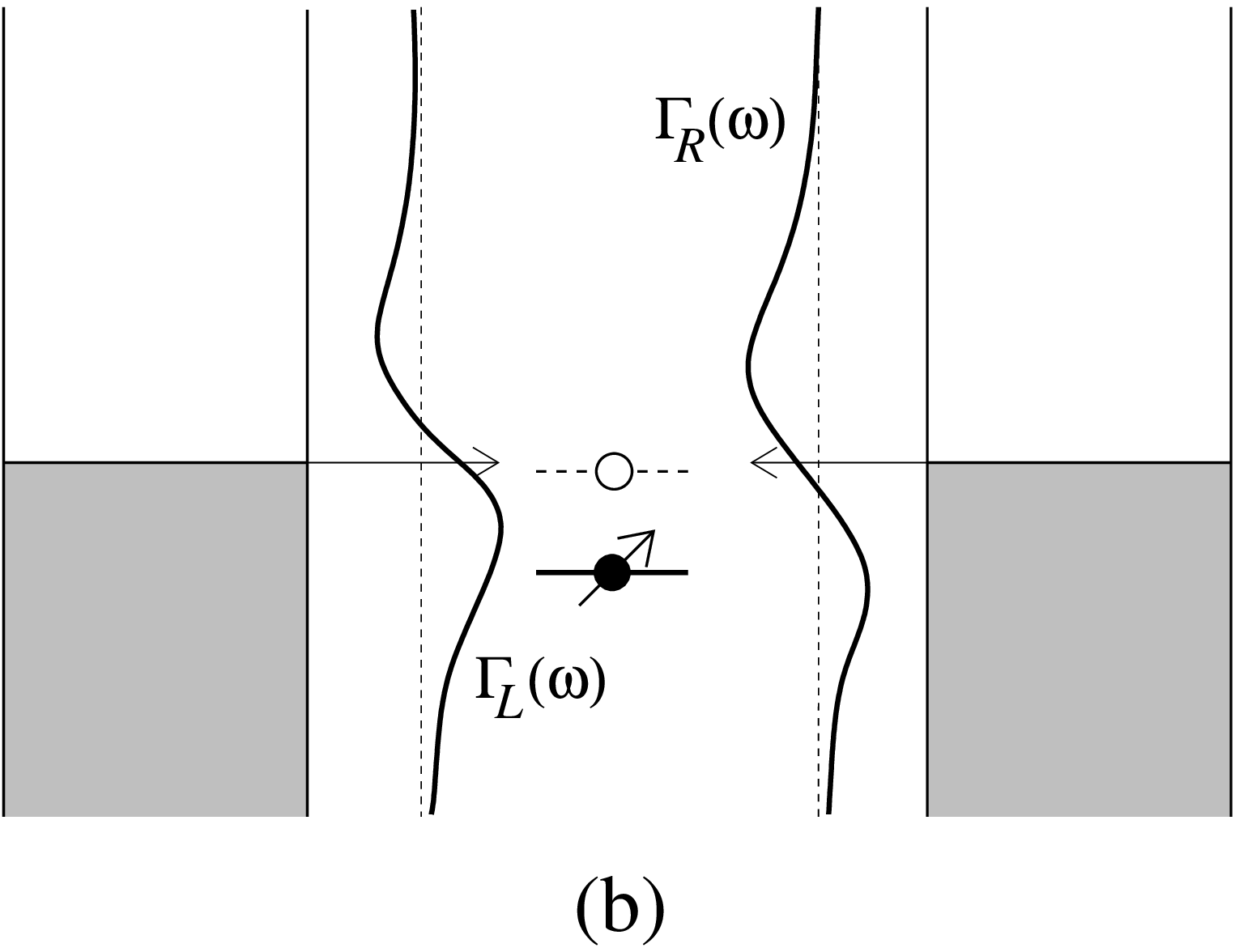}}}
\caption{(a) Schematic energy diagram with chemical potential bias
$\Phi$
applied between the source and drain reservoirs. The empty level
of the quantum dot is aligned at the middle of the two chemical
potentials. (b) After the mapping of steady-state nonequilibrium 
to an effective equilibrium via the boundary condition operator
$\hat{Y}$, the energy diagram of $\hat{H}-\hat{Y}$ has equal
chemical potential. The non-zero current is recovered in the
modified hybridization functions (solid curves) between the QD
and the reservoirs.
}
\label{device}\end{figure}

In this section, we discuss the general
properties of scattering state operators with many-body interactions
present in quantum dot systems and obtain expressions for the bias 
operator in a simple form. We prove that the zero-bias limit in the
mapping of nonequilibrium produces the Kubo formula.
The detailed derivations are presented in Appendix.

The general scattering theory has been well-studied and its
formalism can be found in many textbooks~\cite{bohm}.
However, most of them are studied in terms of state vector
notation and here we present some of the relations in the
operator form.
We first consider a quite general model of quantum dot systems.
The constraint is that the many-body interaction is confined to
the quantum dot.
The Lippmann-Schwinger scattering state operator 
$\psi^\dagger_{\alpha k\sigma}$
is expressed~\cite{gellmann} in the operator equation,
\begin{equation}
\psi^\dagger_{\alpha k\sigma}=c^\dagger_{\alpha k\sigma}
+\frac{1}{\epsilon_{\alpha k}-{\cal L}+i\eta}{\cal L}_V 
c^\dagger_{\alpha k\sigma},
\label{eq:lippmann}
\end{equation}
where $c^\dagger_{\alpha k\sigma}$ is the free continuum
state creation operator inside the reservoir $\alpha$ at
the asymptotic energy $\epsilon_{\alpha k}$ with
the continuum index $k$ and spin $\sigma$. 
$\eta$ is a infinitesimal positive number to define the
asymptotic limit of the scattering state.~\cite{gellmann}
Here the Liouville operators for the full Hamiltonian ${\cal L}$ 
and the interaction ${\cal L}_V$ are defined in the commutation
relation such as ${\cal L}\hat{A}=[\hat{H},\hat{A}]$
for any operator.
The $\alpha(=\pm)$ index refers to the reservoir [$+\leftrightarrow$
source reservoir ($L$), $-\leftrightarrow$ drain reservoir ($R$)].
 
Since the Hamiltonian has many-body interactions, the scattering
state operator $\psi^\dagger$ has terms with many-particle operators,
and it is not clear if they obey the regular fermion anti-commutation
relation. In non-interacting models, it can be shown straightforwardly
that the anti-commutation relation holds. It is shown in Appendix A
that the following commutation relation holds even 
in the interacting limit:
\begin{equation}
\{\psi^\dagger_{\alpha k\sigma},\psi_{\alpha' k'\sigma'}\}
=\delta_{\alpha\alpha'}\delta_{kk'}\delta_{\sigma\sigma'}.
\label{eq:anticommute}
\end{equation}
Therefore, $\psi^\dagger,\psi$ work like typical fermion operators
and any scattering state created by repeated application
of $\psi^\dagger$'s on a particle-vacuum can be considered 
a many-body scattering state.

One of the obstacles in implementing the mapping of nonequilibrium 
has been the lack of understanding of the bias condition operator
suggested by Hershfield~\cite{hershfield},
\begin{equation}
\hat{Y}=\frac{\Phi}{2}\sum_{k\sigma}\left(
\psi^\dagger_{Lk\sigma}\psi_{Lk\sigma}
-\psi^\dagger_{Rk\sigma}\psi_{Rk\sigma}
\right).
\label{eq:y}
\end{equation}
We reduce the $\hat{Y}$ operator in a physically appealing
form in Appendix B as
\begin{equation}
\hat{Y} = \Phi\left[\frac12\sum_{\alpha k\sigma}\alpha c^\dagger_{\alpha k\sigma}
c_{\alpha k\sigma}-\frac{1}{e}\frac{i}{-{\cal L}+i\eta}\hat{I}\right],
\label{eq:ywithi}
\end{equation}
with the current operator $\hat{I}$ which will be defined later.
Although the bias operator has been derived from the boundary
condition imposed with respect to the chemical potential difference, the above
formula relates the potential-driven ensemble to a current-driven 
ensemble~\cite{bokes}.

Expressing the bias operator in terms of current operator
becomes particularly useful in proving the zero-bias limit of 
the conductance. The formulation of nonequilibrium mapping via
$\hat{Y}$ must produce the same Kubo formula in equilibrium
theory. The proof in Appendix C shows that the mapping of nonequilibrium
has the correct description of transport physics in the low-bias
limit. So far, the present formulation of nonequilibrium has been 
shown to be
correct in the non-interacting models at finite bias~\cite{elph}
and in the interacting models in the zero-bias limit.

The notion that we can treat the scattering states as independent
dynamical degrees of freedom has been the central idea in the
mapping of nonequilibrium. In the non-interacting model~\cite{elph}, 
the scattering
state operators are on a firm ground due to the anti-commutation
relation mentioned above and the completeness of the scattering
states encapsulated in the following identity
\begin{equation}
\sum_{\alpha k\sigma}\psi^\dagger_{\alpha k\sigma}\psi_{\alpha k\sigma}
=\sum_{\alpha k\sigma}c^\dagger_{\alpha k\sigma}c_{\alpha k\sigma}
+\sum_\sigma d^\dagger_\sigma d_\sigma,
\end{equation}
provided that there exist no bound states, which is the case
in the limit of large bandwidth.
In the interacting systems, the scattering state
operators $\psi^\dagger_{\alpha k\sigma}$ become dressed with
multi-particle scattering. Therefore, $\psi^\dagger_{\alpha k\sigma}$
should be defined in the many-particle basis and the validity of the
above equation is not obvious. In Appendix D, we derive the relation
for a fairly general class of interactions.

Similarly, behind the idea of writing the bias operator $\hat{Y}$ 
by shifting the chemical potential of $\psi^\dagger_{\alpha k\sigma}$,
we expect for a consistent theory that
\begin{equation}
\hat{H}-\hat{Y}=\sum_{\alpha k\sigma}
\left(\epsilon_{\alpha k\sigma}-\alpha\Phi/2\right)
\psi^\dagger_{\alpha k\sigma}\psi_{\alpha k\sigma},
\end{equation}
and therefore
\begin{equation}
\hat{H}=\sum_{\alpha k\sigma}
\epsilon_{\alpha k\sigma}\psi^\dagger_{\alpha k\sigma}\psi_{\alpha k\sigma}.
\label{eq:complete}
\end{equation}
We derive a general relation for the above summation in Appendix D 
and explicitly demonstrate that the relation Eq.~(\ref{eq:complete}) holds
for the Anderson impurity model. This relation corresponds to the
intertwining relation in the $S$-matrix formalism in the absence of
bound states.
Even if the products $\psi^\dagger_{\alpha k\sigma}\psi_{\alpha k\sigma}$
has a quadratic form, they contain many-body interactions to infinite orders
in the original basis. However, after the summation over all $(\alpha k\sigma)$
indices, only finite order terms up to the original Hamiltonian survive.

\section{Nonequilibrium in Anderson Impurity Model}
\label{sec:anderson}

For the remainder of this paper until Section~\ref{sec:sc}, we will
consider an example of electron transport in the infinite-$U$ Anderson
model as a model for strongly correlated transport system.
In such systems where local interaction dominates
the electronic structure, it is natural to choose the local states
as the unperturbed states. One of the very successful and popular
methods is the slave-boson technique, often used in the limit
of infinite interaction strength. In our model, by letting the
on-site Coulomb interaction to be infinite, we effectively project
out the Hilbert space associated with $d^2$ QD states. Therefore
the available local Hilbert space is $\{|0\rangle, |d_\sigma\rangle:
\sigma=1,\cdots,N\}$, constrained by the relation
\begin{equation}
|0\rangle\langle 0| + \sum_{\sigma=1}^{N}
|d_\sigma\rangle\langle d_\sigma|=1.
\label{eq:constraint}
\end{equation}
Due to the truncation of the Hilbert space, 
the QD states $|d_\sigma\rangle$
cannot be represented by ordinary fermion operators.
One of the techniques to overcome this problem is the slave-boson
method, where we associate the empty state $|0\rangle$ by
a bosonic state $b^\dagger|vac\rangle$, created by a slave-boson
creation operator $b^\dagger$ on an imaginary vacuum $|vac\rangle$
and the singly-occupied states $|d_\sigma\rangle$ by
$f^\dagger_\sigma|vac\rangle$ with pseudo-fermion operator
$f^\dagger_\sigma$. The above constraint can be now recast
in the following operator relation,
\begin{equation}
b^\dagger b + \sum_{\sigma=1}^{N}
f^\dagger_\sigma f_\sigma=1.
\label{eq:project}
\end{equation}
Under the assumption of the above projection, the Hamiltonian
can be written in terms of fermion-boson operators as
\begin{eqnarray}
\hat{H} & = & \hat{H}_0 + \hat{V} \\
\hat{H}_0 & = & \sum_{\alpha k\sigma}\epsilon_{\alpha k}
c^\dagger_{\alpha k\sigma}c_{\alpha k\sigma}
+\epsilon_d\sum_\sigma f^\dagger_\sigma f_\sigma \\
\hat{V} & = & \sum_{\alpha k\sigma}\frac{t_{\alpha k\sigma}}{\sqrt\Omega}
\left(bf^\dagger_\sigma c_{\alpha k\sigma} +
c^\dagger_{\alpha k\sigma}f_\sigma b^\dagger\right),
\end{eqnarray}
with the QD level energy $\epsilon_d$, tunneling amplitude
$t_{\alpha k\sigma}$ and the volume factor $\Omega$.
The unperturbed Hamiltonian $\hat{H}_0$ has the reservoir continua
and local basis as disconnected. The perturbation term $\hat{V}$ turns
on the tunneling between the reservoirs and the quantum dot.

Since the steady-state ensemble must be time-independent, we construct the
nonequilibrium boundary condition in terms of the scattering 
states~\cite{hershfield}.
The scattering state creation operator $\psi^\dagger_{\alpha k\sigma}$
can be written in the Lippmann-Schwinger 
equation~\cite{gellmann,merzbacher,zubarev}:
\begin{equation}
\psi^\dagger_{\alpha k\sigma} = c^\dagger_{\alpha k\sigma}
+\frac{1}{\epsilon_{\alpha k}-{\cal L}_0+i\eta}{\cal L}_V
\psi^\dagger_{\alpha k\sigma},
\label{eq:lippman}
\end{equation}
with the Liouville operator ${\cal L}_0$ with respect to $\hat{H}_0$
defined as ${\cal L}_0\hat{A}=[\hat{H}_0,\hat{A}]$ for an arbitrary
operator $\hat{A}$, and similarly
${\cal L}_V\hat{A}=[\hat{V},\hat{A}]$. 
This formula is equivalent to Eq.~(\ref{eq:lippmann}).
The scattering state operator is expanded in a perturbation series
of the tunneling part ${\cal L}_V$,
\begin{eqnarray}
\psi^\dagger_{\alpha k\sigma} & = &
c^\dagger_{\alpha k\sigma}+\sum_{n=1}^\infty \psi^\dagger_{n,\alpha k\sigma}
\nonumber \\
& = & c^\dagger_{\alpha k\sigma}
+\sum_{n=1}^\infty\left(
\frac{1}{\epsilon_{\alpha k}-{\cal L}_0+i\eta}{\cal L}_V
\right)^n c^\dagger_{\alpha k\sigma}.
\label{eq:psiexp}
\end{eqnarray}
The $n$-th order term can be derived recursively as
\begin{equation}
\psi^\dagger_{n,\alpha k\sigma}=
\frac{1}{\epsilon_{\alpha k}-{\cal L}_0+i\eta}{\cal L}_V 
\psi^\dagger_{n-1,\alpha k\sigma}.
\label{eq:recursion}
\end{equation}
The statistical operator $\hat{Y}$ for the nonequilibrium boundary condition
is defined as in Eq.~(\ref{eq:y}).

If the interaction is local, one can easily see that the scattered
part of the wave-function, $\Delta\psi^\dagger_{\alpha k\sigma}
=\sum_n\psi^\dagger_{n,\alpha k\sigma}$, has the $k$-dependence only through
$\epsilon_{\alpha k}$.
To eliminate unnecessary band-edge effects and bound states, 
we consider a infinite band
Lorentzian density of states (DOS)~\cite{localized} for the both
reservoirs $N_\alpha(\epsilon)=(D/\pi)(\epsilon^2+D^2)^{-1}$ 
with the half-bandwidth $D$.
Then at $\epsilon_{Lk}=\epsilon_{Rk'}$
we have $\Delta\psi^\dagger_{Lk\sigma}=\Delta\psi^\dagger_{Rk'\sigma}$.
After the $k$-summation with the DOS of broad bandwidth and by assuming
the symmetric source-drain parameters, we have a cancellation between
$\Delta\psi^\dagger_{Lk\sigma}\Delta\psi_{Lk\sigma}$ and
$\Delta\psi^\dagger_{Rk\sigma}\Delta\psi_{Rk\sigma}$ in $\hat{Y}$:
\begin{equation}
\hat{Y}
=\hat{Y}_0+
\frac{\Phi}{2}\sum_{k\sigma}\left(
\Delta\psi^\dagger_{Lk\sigma}c_{Lk\sigma}
-\Delta\psi^\dagger_{Rk'\sigma}c_{Rk'\sigma}+h.c.\right),
\label{eq:y1}
\end{equation}
with $\hat{Y}_0=(\Phi/2)\sum_{k\sigma}(
c^\dagger_{Lk\sigma}c_{Lk\sigma}
-c^\dagger_{Rk\sigma}c_{Rk\sigma}).$

\begin{figure}[bt]
\rotatebox{0}{\resizebox{!}{4.2in}{\includegraphics{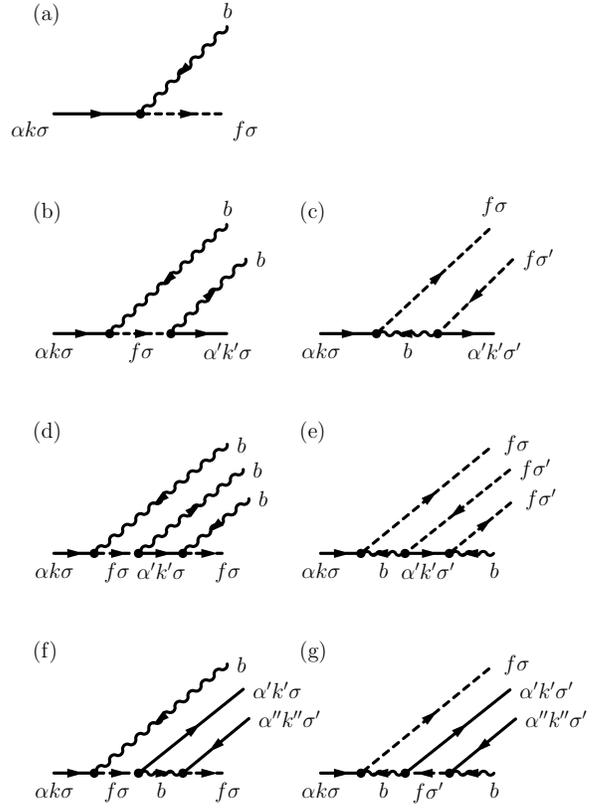}}}
\caption{Diagrams in the expansion of the scattering state
operator $\psi^\dagger_{\alpha k\sigma}$,
expanded perturbatively with respect to the tunneling term $\hat{V}$
in the infinite-$U$ Anderson impurity model.
}
\label{diagram}\end{figure}

\subsection{Expansion of $\psi^\dagger_{\alpha k\sigma}$}

We expand the scattering state operator $\psi^\dagger_{\alpha k\sigma}$
as depicted in FIG.~\ref{diagram}.
FIG.~\ref{diagram}(a) shows the 
diagram for the $n=1$ contribution. The solid line is for the
conduction electron, the dashed line for the pseudo-fermion $f_\sigma$ 
and the wavy line for the slave-boson $b$.
\begin{equation}
\psi^\dagger_{1,\alpha k\sigma}=
\frac{1}{\epsilon_{\alpha k}-{\cal L}_0+i\eta}{\cal L}_V
c^\dagger_{\alpha k\sigma}
=\frac{t_\alpha}{\sqrt\Omega}
\frac{bf^\dagger_\sigma}{\epsilon_{\alpha k}-\epsilon_d+i\eta}.
\label{eq:psi1}
\end{equation}
From now on, we use a shorthand notation
$q\equiv(\alpha k)$. As mentioned above,
we assume the left-right symmetry and the tunneling amplitudes are
independent of $k$, i.e., $t_q=t$. 

In the $n=2$ term, we first compute ${\cal L}_V(bf^\dagger_\sigma)$ as
\begin{equation}
{\cal L}_V(bf^\dagger_\sigma)
=\sum_{q'\sigma'}
\frac{t_{q'}}{\sqrt\Omega}c^\dagger_{q'\sigma'}
(\delta_{\sigma\sigma'}+b^\dagger b\delta_{\sigma\sigma'}
-f_{\sigma'}f^\dagger_\sigma),
\end{equation}
where $b^2=0$ has been used in the above calculation since
any physical state subject to the constraint Eq.~(\ref{eq:project})
is projected out by the application of $b^2$ operator.
Therefore,
\begin{equation}
\psi^\dagger_{2,q\sigma}=
\sum_{q'\sigma'}\frac{t^2_{q'}}{\Omega}
\frac{c^\dagger_{q'\sigma'}
(\delta_{\sigma\sigma'}+b^\dagger b\delta_{\sigma\sigma'}
-f_{\sigma'}f^\dagger_\sigma)}{
(\epsilon_{q}-\epsilon_{q'}+i\eta)(\epsilon_{q}
-\epsilon_d+i\eta)}.
\end{equation}
The second term in the parenthesis of the
numerator corresponds to the diagram
in FIG.~\ref{diagram}(b) and the third term to (c).

The third order term ($n=3$) is obtained from
$\psi^\dagger_{3,q\sigma}=
(\epsilon_{\alpha k}-{\cal L}_0+i\eta)^{-1}{\cal L}_V 
\psi^\dagger_{2,q\sigma}$.
We have simplified the algebra of commutation relations regarding
the QD operators by imposing
the constraint Eq.~(\ref{eq:project}), such as
$bb^\dagger b=b$ and $bf^\dagger_{\sigma'}f_{\sigma'}
f^\dagger_\sigma=bf^\dagger_\sigma
\delta_{\sigma\sigma'}$.
With regards to the conduction electron operators,
we ignore the off-diagonal density and electron (hole) pair
operators as 
\begin{eqnarray}
c_{q''\sigma''}c^\dagger_{q'\sigma'} & \approx &
c_{q'\sigma'}c^\dagger_{q'\sigma'}\delta_{q'q''}\delta_{\sigma'\sigma''}\\
c^\dagger_{q\sigma}c^\dagger_{q'\sigma'},
c_{q\sigma}c_{q'\sigma'} & \approx & 0.
\end{eqnarray}
If we do not consider any effects on transport from
spontaneous magnetic ordering in the continuum
states, {\it i.e.} $\langle c^\dagger_{\alpha k\sigma}
c_{\alpha' k',-\sigma}\rangle=0$, or pairing correlation,
{\it i.e.} $\langle c^\dagger_{\alpha k\sigma}
c^\dagger_{\alpha' k'\sigma'}\rangle=0$, it is reasonable
to drop the corresponding operator products.

Justification of the $q'=q''$ condition seems less
obvious. We discuss this in conjunction with the 
$U=\infty$ condition~\cite{coleman} as follows. As will be discussed 
below, we will replace the operator product 
$c^\dagger_{q'\sigma'}c_{q''\sigma'}$ 
by its expectation value, which amounts to contracting the
out-going electron-hole lines in FIG.~\ref{diagram} (f-g).
However, if $q'\neq q''$ there must be a slave-boson-fermion
loop to mediate the off-diagonal conduction electron states.
This contributes to an extra-charge $Q$ number in the 
Coleman's formulation~\cite{coleman} of the infinite-$U$ model
and can be projected out. It can be argued that the outgoing
conduction electron line will be eventually contracted in 
higher order diagrams. 
However, that corresponds to the so-called
crossing-diagrams~\cite{nca} which are of higher order in $1/N$ and
negligible in the large $N$ limit.
Therefore, in the context of the large-$U$ and large-$N$ limit, 
the off-diagonal density operators are ignored.

With above approximations,
the third order contribution of the scattering state operator
becomes
\begin{equation} 
\psi^\dagger_{3,q\sigma} =
 \frac{t_\alpha/\sqrt\Omega}{(\epsilon_{q}-\epsilon_d+i\eta)^2}
\left(-i\Gamma+\hat{\Sigma}_r(\epsilon_q)\right)
bf^\dagger_\sigma,
\end{equation}
where $\Gamma=\Gamma_L+\Gamma_R,\ \Gamma_\alpha=\Omega^{-1}
\pi t^2\sum_{k}\delta(\epsilon-\epsilon_{\alpha k})=
\pi t^2 N(0)$ with $N(0)$ the density of states of the reservoirs.
The retarded self-energy operator is defined as
\begin{equation}
\hat{\Sigma}_r(\epsilon)=\frac{t_\alpha^2}{\Omega}
\sum_{q',\sigma'\neq\sigma}\frac{c^\dagger_{q'\sigma'}c_{q'\sigma'}}{
\epsilon_{q}-\epsilon_{q'}+i\eta}.
\end{equation}
Applying the same approximations discussed above, we obtain higher
order perturbation terms in a geometric series as
\begin{eqnarray}
\psi^\dagger_{{\rm odd},q\sigma} & = &
\sum_{n=1,3,\cdots,\infty}\psi^\dagger_{n,q\sigma} \\
& = & \frac{t_\alpha}{\sqrt\Omega}bf^\dagger_\sigma\left(
\epsilon_{\alpha k}-\epsilon_d+i\Gamma
-\hat{\Sigma}_r(\epsilon_q) \right)^{-1}.
\label{eq:psiodd}
\end{eqnarray}
So far the self-energy {\it operator} is derived from the
equation of motion and is independent of the boundary condition.

Here we replace the pairs of electron
operators in the numerator of the self-energy operator with average
values in the asymptotic limit,
$c^\dagger_{q,\sigma}c_{q\sigma}\approx\langle
\psi^\dagger_{\alpha k\sigma}\psi_{\alpha k\sigma}\rangle=
f(\epsilon_{\alpha  k}-\alpha\Phi/2)\equiv 
f_\alpha(\epsilon_{\alpha  k})$, with the Fermi-Dirac
function $f(\epsilon)=(1+e^{\beta\epsilon})^{-1}$.
The different chemical 
potential of the source-drain reservoirs is taken into account 
in the $\alpha\Phi/2$ term. 
By taking the expectation value over the self-energy operator,
we have incorporated the boundary condition
in the scattering state. We define an effective 
retarded Green function $g_d(\epsilon)=\left[
\epsilon_{\alpha k}-\epsilon_d+i\Gamma
-\Sigma^r(\epsilon_q)\right]^{-1}$
in terms of the retarded self-energy $\Sigma^r(\epsilon)$ given as
\begin{eqnarray}
\Sigma^r(\epsilon_q) & = & \frac{1}{\Omega}\sum_{q'\sigma'\neq\sigma}
\frac{t_\alpha^2 f_{\alpha'}(\epsilon_{q'})}{
\epsilon_{q}-\epsilon_{q'}+i\eta} \nonumber \\
& = & (N-1)\sum_{\alpha'}\frac{\Gamma_{\alpha'}}{\pi}
\int d\epsilon' \frac{n_{\alpha'}(\epsilon')
f_{\alpha'}(\epsilon')}{\epsilon_{q}-\epsilon'+i\eta},
\label{eq:retself}
\end{eqnarray}
with the renormalized DOS for the $\alpha$-th reservoir
$n_{\alpha'}(\epsilon')=N_{\alpha'}(\epsilon')/N_{\alpha'}(0)$.
The self-energy $\Sigma^r(\epsilon)$ has a logarithmic singularity
in the limit of $T=0$ for $\epsilon_q\rightarrow 
\alpha \Phi/2$ as
\begin{eqnarray}
& & 
\int d\epsilon' \frac{n_{\alpha'}(\epsilon')
f_{\alpha'}(\epsilon')}{\epsilon_{q}-\epsilon'+i\eta} \nonumber \\
& \approx & \ln\frac{1}{|\epsilon_q-\alpha \Phi/2|}
-i\pi n_{\alpha'}(\epsilon_q)\theta
\left(\alpha \Phi/2-\epsilon_q\right).
\label{eq:logarithm}
\end{eqnarray}
This has the familiar form of the Brillouine-Wigner perturbation
theory~\cite{nca} and goes back to the typical expression 
in the zero-bias limit (with a single chemical potential).
We now write the odd-order perturbation terms of the 
scattering state operator as
\begin{equation}
\Delta\psi^\dagger_{{\rm odd},q\sigma}=
\frac{t}{\sqrt\Omega}g_d(\epsilon_q)
bf^\dagger_\sigma,
\label{eq:psioddg}
\end{equation}
which has the same form as in the non-interacting resonant
level QD system when $g_d(\epsilon_q)$ is replaced with 
the non-interacting function~\cite{elph} 
$(\epsilon_{\alpha k}-\epsilon_d+i\Gamma)^{-1}$.
The renormalization factor due to $\Sigma^r(\epsilon)$ is
essential to construct the bias operator $\hat{Y}$
in the strongly correlated limit. As will be shown later,
anomalous Kondo peak at zero bias cannot be produced
without this logarithmic singularity.

The approximations used here are motivated to produce
physically intuitive formula, similar to the scattering 
states in the non-interacting limit. We view that the
many-body particle terms in the scattering state operators
as out-going single particle states dressed with contractible 
many-particle excitations. One can interpret
the replacement of many-particle operators by single-particle
operators with renormalized amplitude as a quasi-particle
approximation, in the similar spirit of the Fermi-liquid theory.

The even order perturbation terms can be obtained from
$\Delta\psi^\dagger_{{\rm even},q\sigma}=(\epsilon_q-{\cal L}_0+i\eta)^{-1}
{\cal L}_V\Delta\psi^\dagger_{{\rm odd},q\sigma}$ and, using 
the approximations leading to Eq.~(\ref{eq:psioddg}), we get
\begin{equation}
\Delta\psi^\dagger_{{\rm even},q\sigma}=
\frac{t^2}{\Omega}\sum_{q'\sigma'}
\frac{g_d(\epsilon_q)}{\epsilon_{q}-\epsilon_{q'}+i\eta}
c^\dagger_{q'\sigma'}
(\delta_{\sigma\sigma'}+b^\dagger b\delta_{\sigma\sigma'}
-f_{\sigma'}f^\dagger_\sigma),
\end{equation}
which can be interpreted as the renormalized scattered wave.
Finally the approximate scattering state operator is
\begin{eqnarray}
\psi^\dagger_{q\sigma} & = & c^\dagger_{q\sigma}
+\frac{t}{\sqrt\Omega}g_d(\epsilon_q)
bf^\dagger_\sigma \\
\label{eq:psitotal}
& & +\frac{t^2}{\Omega}\sum_{q'\sigma'}
\frac{g_d(\epsilon_q)}{\epsilon_{q}-\epsilon_{q'}+i\eta}
c^\dagger_{q'\sigma'}
(\delta_{\sigma\sigma'}+b^\dagger b\delta_{\sigma\sigma'}
-f_{\sigma'}f^\dagger_\sigma), \nonumber
\end{eqnarray}
which has the similar structure of the non-interacting scattering state.

The physical meaning of the above scattering state is quite transparent.
The second term is for the conduction electrons
to tunnel on to the QD site with the amplitude given by the full Green
function. 
The last term involving continuum states is either 
from the potential scattering or
from the exchange of the conduction with
the QD electrons. Due to the approximation of $U=\infty$, local 
electrons first tunnel out to the reservoir
and the empty local state is subsequently filled
by a conduction electron from the reservoir
($d^1\to d^0\to d^1$).

If the outgoing conduction electron were to be in the same spin state
as the incoming spin, there are two possibilities.
First, when the initial QD is in the empty state
$b^\dagger|vac\rangle$, a conduction electron goes through the
potential scattering to hop onto
the QD and then tunnels out with the same spin
($d^0\to d^1\to d^0$). It can be seen explicitly from 
Eq.~(\ref{eq:psitotal}) that the initially empty QD state
($b^\dagger|vac\rangle$) does not change after the scattering
[$(\delta_{\sigma\sigma'}+b^\dagger b\delta_{\sigma\sigma'}
-f_{\sigma'}f^\dagger_\sigma)
b^\dagger|vac\rangle=\delta_{\sigma\sigma'}b^\dagger|vac\rangle$].
Second, if the QD was singly occupied with the same spin
as the incoming state
($f^\dagger_\sigma|vac\rangle$), the QD electron exchanges with
the incoming electron without flipping the spin,
[$(1+b^\dagger b-f_{\sigma}f^\dagger_\sigma) f^\dagger_\sigma
|vac\rangle=f^\dagger_\sigma|vac\rangle$].
Therefore, when the incoming and out-going waves have the same spin,
the factor $(\delta_{\sigma\sigma'}+b^\dagger b\delta_{\sigma\sigma'}
-f_{\sigma'}f^\dagger_\sigma)$ can be considered as one.

However, if the QD had a different spin state 
($f^\dagger_{\sigma'}|vac\rangle$, $\sigma'\neq\sigma$) from
the incoming electron, 
the outgoing electron cannot be
in the same spin state as the incoming one
[$(1+b^\dagger b-f_{\sigma}f^\dagger_\sigma) f^\dagger_{\sigma'}
|vac\rangle=0$], since $d^2$-configurations are forbidden by
the $U=\infty$ condition and therefore 
the QD electron must hop out first in a scattering event.

\subsection{Effective Hamiltonian and Non-crossing Approximation}

Here, we focus
on how the reservoir-QD tunneling is modified by the bias.
The modified tunneling results from the
second term in Eq.~(\ref{eq:psitotal}), and the 
bias operator Eq.~(\ref{eq:y1}) is written as
\begin{eqnarray}
\hat{Y} & = & \frac{\Phi}{2}\sum_{\alpha k}\alpha\left[
c^\dagger_{\alpha k\sigma}c_{\alpha k\sigma}
+\frac{t}{\sqrt\Omega}g_d(\epsilon_{\alpha k})
bf^\dagger_\sigma c_{\alpha k\sigma} + h.c.
\right] \nonumber \\
& & +\hat{Y}_{cc},
\end{eqnarray}
where the last term $\hat{Y}_{cc}$ accounts for the non-diagonal
coupling between conduction electrons, proportional to
$c^\dagger_{\alpha k\sigma}c_{\alpha' k'\sigma'}$, created by the
steady-state boundary condition.
Finally an approximate
effective nonequilibrium Hamiltonian $\hat{\cal H}=\hat{H}-\hat{Y}$
becomes
\begin{eqnarray}
\hat{\cal H} & = & \hat{H}-\hat{Y} \nonumber \\
& = & \sum_{\alpha k\sigma}\left[\epsilon_{\alpha k}
-\frac{\alpha\Phi}{2}\right]
c^\dagger_{\alpha k\sigma}c_{\alpha k\sigma}
+\epsilon_d\sum_\sigma f^\dagger_\sigma f_\sigma 
\label{eq:hamil} \\
& + & \frac{t}{\sqrt\Omega}\sum_{\alpha k\sigma}\left[\left(
1-\frac{\alpha\Phi}{2} g_d(\epsilon_{\alpha k})\right)
bf^\dagger_\sigma c_{\alpha k\sigma} + h.c.\right]
-\hat{Y}_{cc}.\nonumber
\end{eqnarray}

Since the scattering state Eq.~(\ref{eq:psitotal}) is an extended
state, the product of $\psi^\dagger_{q\sigma}\psi_{q\sigma}$
in $\hat{Y}$ produces non-local interaction between all possible
basis states and Hamiltonians with only local interaction become
non-local after the mapping. Usually we need a truncation scheme to
treat the non-locality~\cite{elph}. In the above Hamiltonian,
the non-local effects are implicitly included in the modified
tunneling and we do not consider additional non-local effects
in this work.

In this section, we calculate the the on-site QD Green function 
${\cal G}_d(\omega)$. Since the non-local terms in $Y_{cc}$
have indirect effects to ${\cal G}_d(\omega)$, we temporarily
drop $Y_{cc}$ in the calculation of ${\cal G}_d(\omega)$.
We use curly symbols, such as
$\hat{\cal H}$ and ${\cal G}_d(\omega)$, to denote the quantities
which are derived with $\hat{H}-\hat{Y}$ as the time-evolution
generator.

To solve the effective Hamiltonian subject to the constraint
condition Eq.~(\ref{eq:project}), we use the Coleman's
projection method~\cite{coleman} and perform the nonperturbative
partial summation of Feynman diagrams in the non-crossing
approximation (NCA)~\cite{nca}. The constraint condition
is imposed within the grand-canonical ensemble scheme
with respect to the quantum dot charge operator 
$\hat{Q}$ defined as
$\hat{Q}=b^\dagger b+\sum_\sigma f^\dagger_\sigma f_\sigma$.
Only the subspace $Q_1$ (with $\hat{Q}=1$)
is physically meaningful. The grand-canonical ensemble 
is introduced with a fictitious chemical 
potential $-\lambda$ associated with $\hat{Q}$ in the
partition function $Z_G(\lambda)$ as
\begin{equation}
Z_G(\lambda)={\rm Tr}\,e^{-\beta(\hat{\cal H}+\lambda \hat{Q})}
=\sum_{Q=0}^\infty Z_c(Q)e^{-\beta\lambda Q},
\end{equation}
with the partition function $Z_c(Q)$ in the
subspace of the fixed charge $\hat{Q}=Q$. The expectation value of an operator
$\hat{A}$ averaged over $Z_c(Q=1)$ is accomplished by taking the
$\lambda\rightarrow\infty$ limit as
\begin{equation}
\langle\hat{A}\rangle=\lim_{\lambda\rightarrow\infty}
\frac{\langle \hat{A}\hat{Q}\rangle_\lambda}{
\langle\hat{Q}\rangle_\lambda},
\end{equation}
where the average $\langle \cdots\rangle_\lambda$ is taken
over the grand-canonical ensemble before $\lambda$
is taken to the infinity and regular Feynman diagram technique
can be applied to evaluate the average.

\begin{figure}[bt]
\rotatebox{0}{\resizebox{3.0in}{!}{\includegraphics{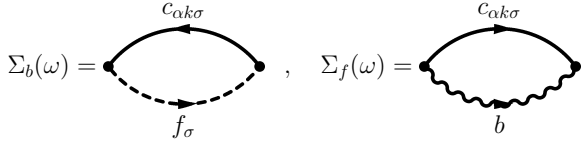}}}
\caption{Non-crossing approximation diagrams for the
self-energies of slave-boson ($b$) and pseudo-fermion 
($f_\sigma$).
}
\label{fig:nca}\end{figure}

Feynman diagrams are partially summed by treating the inverse of
the QD degeneracy, $1/N$, as the expansion parameter. The
non-crossing approximation exploits the fact that
diagrams with crossing one-particle propagators contribute
to the high order expansion in $O(1/N^2)$, and therefore
can be ignored in the limit of large QD degeneracy.
This procedure can be summed up in the self-energy diagrams depicted
in FIG.~\ref{fig:nca}. The dashed line represents
the {\it full} Green function for the pseudo-fermion,
the wavy line for the slave-boson, and the solid line
for the bare conduction electron. After $\lambda\rightarrow
\infty$ limit is taken, the self-energies 
$\Sigma_b(\omega),\Sigma_f(\omega)$ for $b,f_\sigma$ can
be expressed as
\begin{eqnarray}
\Sigma_b(\omega) & = & N\sum_{\alpha=\pm}
\int d\epsilon \frac{\tilde{\Gamma}_\alpha(\epsilon)}{\pi}
f_\alpha(\epsilon)G_f(\omega+\epsilon'_\alpha) \label{eq:selfb}\\
\Sigma_f(\omega) & = & \sum_{\alpha=\pm}
\int d\epsilon \frac{\tilde{\Gamma}_\alpha(\epsilon)}{\pi}
[1-f_\alpha(\epsilon)]G_b(\omega-\epsilon'_\alpha),
\label{eq:selff}
\end{eqnarray}
with $\epsilon'_\alpha=\epsilon-\alpha\Phi/2$.
$G_b(\omega),G_f(\omega)$ are full one-particle Green functions
for the slave-boson and pseudo-fermion, respectively, defined as
\begin{eqnarray}
G_b(\omega) & = & [\omega-\Sigma_b(\omega)]^{-1}\\
G_f(\omega) & = & [\omega-\epsilon_d-\Sigma_f(\omega)]^{-1}.
\end{eqnarray}
The effective hybridization function $\tilde{\Gamma}_\alpha(\epsilon)$
is defined with the modified tunneling amplitude in Eq.~(\ref{eq:hamil})
as
\begin{equation}
\tilde{\Gamma}_\alpha(\epsilon) = \pi t^2\left|
1-\frac{\alpha\Phi}{2} g_d(\epsilon)
\right|^2 N_\alpha(\epsilon).
\label{eq:hyb}
\end{equation}
The closed set of Eqs.~(\ref{eq:selfb}-\ref{eq:hyb}) is 
self-consistently solved.

The shifted continuum energy levels and the effective hybridization
function in $\hat{\cal H}$ is schematically sketched in 
FIG.~\ref{device}(b). The dashed line is the original hybridization 
function at zero bias with a flat DOS. As the bias $\Phi$ is turned on,
the hybridization is modified by the term proportional to
$\Phi g_d(\epsilon)$ in Eq.~(\ref{eq:hyb}), as drawn in solid line.
The enhancement of the effective hybridization from the source 
reservoir $\Gamma_L(\omega)$ at lower energy can be understood as follows.
The bias voltage creates a current-carrying nonequilibrium ensemble.
However, after adjusting the chemical potentials of the two reservoirs
to the same level, the electron current ($L\to R$) should be
restored via the effective tunneling. By enhancing the tunneling magnitude,
hence the hybridization,
to the source reservoir and suppressing the drain-hybridization
for the low energy states, we recover the electron flow from $L$ to $R$.
Similarly, the enhanced drain-hybridization at higher energy promotes
the hole-current from $R$ to $L$. In fact, it is not only the magnitude
of the tunneling parameters but also the phase factor (coming from the
retarded Green function) associated with the tunneling electrons
which contribute to the current. 
This phase factor breaks the time-reversal symmetry.

The on-site Green function of the QD state,
${\cal G}_d(\omega)$,
can be expressed as a convoluted integral of $b-f$ spectral functions
as
\begin{equation}
{\cal G}_d(\omega)=\int d\epsilon\frac{\rho_d(\epsilon)}{\omega-\epsilon+i\eta},
\label{eq:gd}
\end{equation}
with its spectral function $\rho_d(\epsilon)$ given as
\begin{equation}
\rho_d(\epsilon)=\frac{1+e^{-\beta\epsilon}}{Z_f}\int d\epsilon'
\rho_b(\epsilon')\rho_f(\epsilon+\epsilon'),
\end{equation}
where $\rho_b(\epsilon),\rho_f(\epsilon)$ are spectral functions of
the $b$, $f$ operators, respectively, and the local partition
function $Z_f$ is given as
\begin{equation}
Z_f=\int d\epsilon\,e^{-\beta\epsilon}\left[\rho_b(\epsilon)
+N\rho_f(\epsilon) \right].
\end{equation}

\subsection{Current-voltage relation}

\begin{figure}[bt]
\rotatebox{0}{\resizebox{3.0in}{!}{\includegraphics{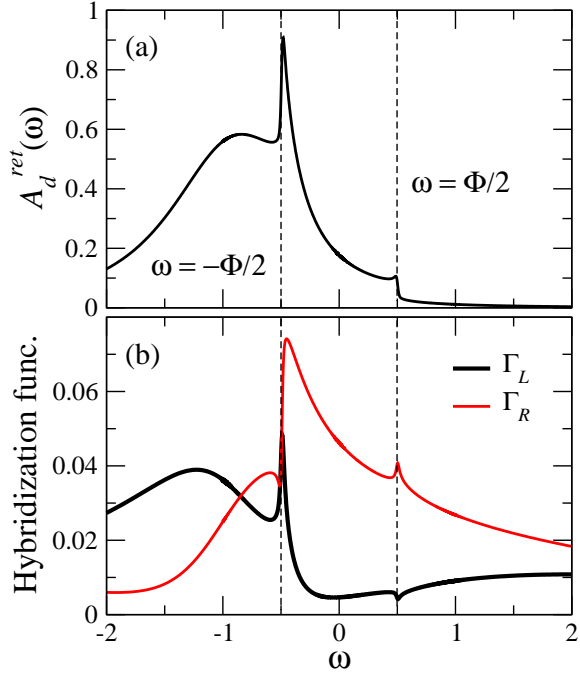}}}
\caption{Spectral function of the retarded Green function 
for the quantum dot site and
the hybridization function from the left- and right-reservoir.
The bias $\Phi$ is set to 1, $t_L=t_R=0.48$ and $\beta=1/T=160$.
The Fermi energy positions for the left ($\omega=\Phi/2$)
and the right ($\omega=-\Phi/2$) reservoirs are marked by
dashed lines.
}
\label{fig:hyb}\end{figure}

The current operator can be derived from the continuity equation
$\partial (ed^\dagger_\sigma d_\sigma)/\partial t
+\hat{I}_{L\sigma}+\hat{I}_{R\sigma}=0$. The current
operator through the left- and right-reservoirs are obtained as
\begin{eqnarray}
\hat{I}_{\alpha\sigma} & = & 
\frac{iet}{\sqrt\Omega}\sum_k(c^\dagger_{\alpha k\sigma}
d_\sigma-d^\dagger_\sigma c_{\alpha k\sigma}) \nonumber \\
& = & \frac{iet}{\sqrt\Omega}\sum_k(c^\dagger_{\alpha k\sigma}
b^\dagger f_\sigma-bf^\dagger\sigma c_{\alpha k\sigma}).
\label{eq:curr}
\end{eqnarray}
In the steady-state condition, $\langle d^\dagger_\sigma d_\sigma\rangle$ 
is constant and 
\begin{equation}
I_L+I_R=0
\label{eq:isym}
\end{equation}
with $I_\alpha=\sum_\sigma\langle \hat{I}_{\alpha\sigma}\rangle$. 
If we ignore the $\hat{Y}_{cc}$ term in the effective Hamiltonian
as in the previous section, we violate the symmetry
condition Eq.~(\ref{eq:isym}) as follows.
Without the $\hat{Y}_{cc}$ term in $\hat{\cal H}$,
the expectation value of operators appearing in Eq.~(\ref{eq:curr})
can be expressed in the finite-temperature Green functions as
\begin{equation}
\frac{it}{\sqrt\Omega}\sum_k
\langle bf^\dagger_\sigma c_{\alpha k\sigma}\rangle
=it^2\frac{1}{\beta}\sum_{\omega_n}\frac{1}{\Omega}\sum_k
\frac{{\cal G}_d(i\omega_n)}{i\omega_n-\epsilon_{\alpha k}+\alpha \Phi/2},
\label{eq:Iwrong}
\end{equation}
where the QD-Green function Eq.~(\ref{eq:gd}) has been analytically
continued to the imaginary Matsubara frequencies, 
$i\omega_n=i(2n+1)\pi/\beta$ at integer $n$.
The on-site Green function ${\cal G}_d(i\omega_n)$ is independent of the
reservoir index $\alpha$. As discussed in the previous section, 
the tunneling from the source
reservoir is dominated by the electron-hopping, $d^1\rightarrow d^2$,
while the tunneling from the drain
reservoir is dominated by the hole-hopping, $d^1\rightarrow d^0$.
Therefore, the modified tunneling in the source and drain-hybridization
do not have the same values of at a given energy (See FIG.~\ref{fig:hyb})
and the expectation
value computed from Eq.~(\ref{eq:Iwrong}) does not satisfy the
charge conservation, Eq.~(\ref{eq:isym}).

Physically, it is no surprise that the $\hat{Y}_{cc}$ term is crucial
for the current. $\hat{Y}_{cc}$ has the direct coupling between 
the source and drain reservoirs driven by the bias and has the
effect of closing the circuit. Furthermore, phase factors in the
coupling act as {\it current battery} which forces a net current from
the source to drain through the quantum dot.

To restore the $L-R$ symmetry, off-diagonal Green functions for 
conduction electrons $G^0_{qq'}(i\omega_n)$, resulting from
$\hat{Y}_{cc}$,
has to be considered in Eq.~(\ref{eq:Iwrong}),
\begin{eqnarray}
& & \frac{it}{\sqrt\Omega}\sum_k
\langle bf^\dagger_\sigma c_{\alpha k\sigma}\rangle \nonumber \\
& = & 
it^2\frac{1}{\beta}\sum_{\omega_n}\frac{1}{\Omega}\sum_{qq'}
G^0_{qq'}(i\omega_n){\cal G}_d(i\omega_n).
\label{eq:Icorrect}
\end{eqnarray}
Here the conduction electron Green function 
$G^0_{qq'}(i\omega_n)$ is a propagator
in the so-called `cavity' Hamiltonian~\cite{dmft}, where
the QD site and its tunneling to the
reservoirs are removed from the full Hamiltonian $\hat{\cal H}$.
It should be emphasized that since
the bias operator $\hat{Y}$ is
built upon the correlated scattering states, even the `cavity' 
part of the 
Hamiltonian has correlation effects.

Recalling that much of the correlations effects 
are already present in the full Green function $g_d(\epsilon)$
in the coefficient of the scattered wave [third term in 
Eq.~(\ref{eq:psitotal})], we approximate
Eq~(\ref{eq:psitotal}), in analogy to the
non-interacting scattering state, as
\begin{equation}
\tilde{\psi}^\dagger_{q\sigma}=c^\dagger_{q\sigma}
+\frac{t_\alpha}{\sqrt\Omega}g_d(\epsilon_q)
d^\dagger_\sigma
+\frac{1}{\Omega}\sum_{q'}
\frac{t_\alpha t_{\alpha'}g_d(\epsilon_q)
}{\epsilon_{q}-\epsilon_{q'}+i\eta}
c^\dagger_{q'\sigma},
\label{eq:psiapprox}
\end{equation}
which amount to replacing the factor $(\delta_{\sigma\sigma'}+
b^\dagger b\delta_{\sigma\sigma'}-f_{\sigma'}f^\dagger_\sigma)$ 
by $\delta_{\sigma\sigma'}$.
As discussed at the end of Section~\ref{sec:anderson}-A, 
$(\delta_{\sigma\sigma'}+
b^\dagger b\delta_{\sigma\sigma'}-f_{\sigma'}f^\dagger_\sigma)
=\delta_{\sigma\sigma'}$ without spin-flip scattering.

\begin{figure}[bt]
\rotatebox{0}{\resizebox{1.2in}{!}{\includegraphics{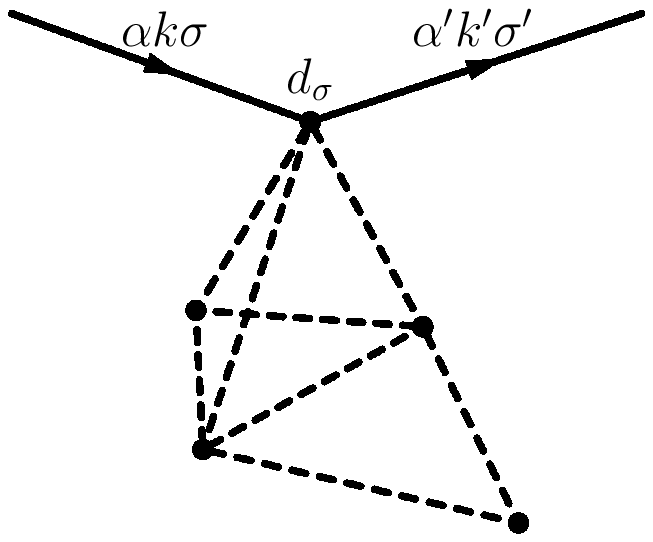}}}
\rotatebox{0}{\resizebox{1.6in}{!}{\includegraphics{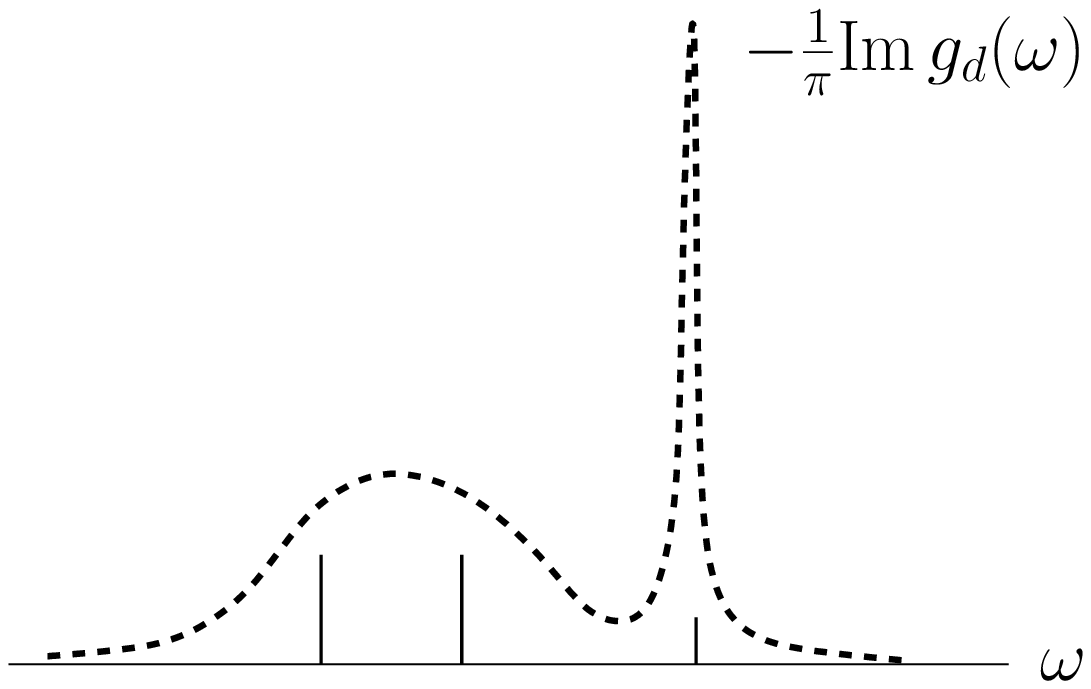}}}
\caption{Coupling between the continuum states in $\hat{H}-\hat{Y}$
is reconstructed by introducing a QD cluster and fitting the
QD Green function $g_d(\omega)$. 
}
\label{fig:geff}\end{figure}

We now obtain a simple expression for the off-diagonal Green 
function $G^0_{qq'}(i\omega_n)$ over the
`cavity' Hamiltonian based on the non-interacting scattering state
operators $\tilde{\psi}^\dagger_{q\sigma}$ using the following trick.
The Green function $g_d(\epsilon)$ is fitted by introducing
an arbitrary, but finite~\cite{finite}
number of discrete levels which couple
to the QD site [see FIG.~\ref{fig:geff}(a)]. 
For example, we
prepare a finite cluster QD of arbitrary geometry, with the 
resulting energy poles distributed 
under the envelope function given by the spectral function
of $g_d(\omega)$, [$=-\pi^{-1}{\rm Im}g_d(\omega)$],
as depicted in FIG.~\ref{fig:geff}(b).
We choose the finite cluster such that the peak positions
and their weight match the quasi-particle energies and
the renormalization factor.
Broad peaks can be fitted
by introducing multiple cluster peaks underneath, as sketched
in FIG.~\ref{fig:geff}(b). The following derivation does not
depend on a particular fitting scheme as long as the fitting
can be made accurate.

Then, the effective Hamiltonian
\begin{equation}
\tilde{H}-\tilde{Y}=\sum_{\alpha k\sigma}
\left[\epsilon_{\alpha k\sigma}-\frac{\alpha\Phi}{2}\right]
\tilde{\psi}^\dagger_{q\sigma}\tilde{\psi}^\dagger_{q\sigma},
\label{eq:htilde}
\end{equation}
can approximate the `cavity' part $\hat{Y}_{cc}$
in the original Hamiltonian Eq.~(\ref{eq:hamil}).
Although the QD parts of $\tilde{H}-\tilde{Y}$ and $\hat{H}-\hat{Y}$ can be
quite different, it does not matter as long as we are only concerned
with the `cavity' Green functions.
Carrying out the similar calculations as detailed
in the Appendix of Ref.~\cite{elph}, the $c-d$
Green function defined as
\begin{eqnarray}
& &  \tilde{G}_{Ld}(i\omega_n) \\
& = &
\left\langle c_{L\sigma}\frac{1}{i\omega_n-{\cal L}_{H-Y}}d^\dagger
\right\rangle
+\left\langle d^\dagger \frac{1}{i\omega_n+{\cal L}_{H-Y}}c_{L\sigma}\right\rangle,
\nonumber
\end{eqnarray}
with $c_{L\sigma}=\sqrt{\Omega^{-1}}\sum_k c_{Lk\sigma}$,
is given by
\begin{eqnarray}
& & t_L\left[\tilde{G}_{Ld}(i\omega_n)-\tilde{G}_{dL}(i\omega_n)\right]
\nonumber \\
& = &
2i\frac{\Gamma_L\Gamma_R}{\Gamma}\left[
g_d(i\omega_n+\Phi/2)-g_d(i\omega_n-\Phi/2)\right].
\end{eqnarray}
The symmetry between the left and right is now evident.
Similarly, the QD Green function can be expressed as
\begin{equation}
\tilde{G}_d(i\omega_n)=\frac12\left[
g_d(i\omega_n+\Phi/2)+g_d(i\omega_n-\Phi/2)\right].
\end{equation}
The `cavity' Green function then satisfies the Dyson
equation
\begin{eqnarray}
& & t_L\left[\tilde{G}_{Ld}(i\omega_n)-\tilde{G}_{dL}(i\omega_n)\right]
\nonumber \\
& = &
t_L t_R\left[G^0_{LR}(i\omega_n)-G^0_{RL}(i\omega_n)\right]
\tilde{G}_d(i\omega_n) \nonumber \\
& = & 
4i\frac{\Gamma_L\Gamma_R}{\Gamma}
\frac{g_d(i\omega_n+\Phi/2)-g_d(i\omega_n-\Phi/2)}{
g_d(i\omega_n+\Phi/2)+g_d(i\omega_n-\Phi/2)}
\tilde{G}_d(i\omega_n)\nonumber
\label{eq:dyson0}
\end{eqnarray}
Finally, 
after replacing $\tilde{G}_d(i\omega_n)$ in
the above equation by the
nonequilibrium Green function ${\cal G}_d(i\omega_n)$, analytically
continued from Eq.~(\ref{eq:gd}),
the Dyson equation for the full off-diagonal
Green functions $G_{Ld}(i\omega_n)$ and $G_{dL}(i\omega_n)$
becomes
\begin{eqnarray}
& & t_L\left[G_{Ld}(i\omega_n)-G_{dL}(i\omega_n)\right] \\
& = &
4i\frac{\Gamma_L\Gamma_R}{\Gamma}
\frac{g_d(i\omega_n+\Phi/2)-g_d(i\omega_n-\Phi/2)}{
g_d(i\omega_n+\Phi/2)+g_d(i\omega_n-\Phi/2)}{\cal G}_d(i\omega_n).
\nonumber
\end{eqnarray}
From $\langle c^\dagger_{L\sigma}d_\sigma\rangle=-\beta^{-1}\sum_n
G_{dL}(i\omega_n)$, the current can be expressed as
\begin{eqnarray}
& & I(\Phi) 
 = N\frac{ie}{\hbar}t_L[\langle d^\dagger_\sigma c_{L\sigma} \rangle
-\langle c^\dagger_{L\sigma} d_\sigma\rangle] 
\label{eq:currfinal} \\
& = & \frac{4Ne}{\hbar}\frac{\Gamma_L\Gamma_R}{\Gamma\beta}
\sum_n
\frac{g_d(i\omega_n+\Phi/2)-g_d(i\omega_n-\Phi/2)}{
g_d(i\omega_n+\Phi/2)+g_d(i\omega_n-\Phi/2)}{\cal G}_d(i\omega_n).
\nonumber
\end{eqnarray}
The numerator $g_d(i\omega_n+\Phi/2)-g_d(i\omega_n-\Phi/2)$
acts as an energy window opened between the source-drain
chemical potential $\pm\Phi/2$ and the Matsubara summation performs
the integration over the window.
This expression becomes exact~\cite{elph} in the non-interacting 
limit with the
non-interacting retarded Green function $g^0_d(z)$ as
\begin{equation}
I_0(\Phi) =\frac{4Ne}{\hbar}\frac{\Gamma_L\Gamma_R}{\Gamma}
\frac1\beta\sum_n
\frac12[g^0_d(i\omega_n+\Phi/2)-g^0_d(i\omega_n-\Phi/2)].
\label{eq:curr0}
\end{equation}

\section{Results}
\label{sec:results}

\begin{figure}[bt]
\rotatebox{0}{\resizebox{3.0in}{!}{\includegraphics{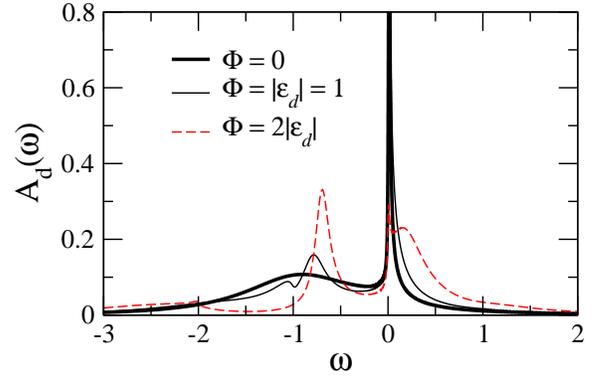}}}
\caption{Spectral function of the thermal Green function 
with $\hat{H}-\hat{Y}$ as the time-evolution generator.
The bias $\Phi$ is set to 1, $t_L=t_R=0.48$ and $\beta=1/T=160$.
}
\label{fig:rho}\end{figure}

We first discuss the retarded Green function of the quantum
dot, $g_d(\epsilon)$,
which appear in the expression for the scattering state
operator $\psi^\dagger_{\alpha k\sigma}$, Eq.~(\ref{eq:psitotal}).
FIG.~\ref{fig:hyb}(a) shows the spectral function, $-\pi^{-1}{\rm Im}
g_d(\omega)$, at the unit bias $\Phi=1$ with tunneling parameters
$t_L=t_R=0.48$ and the inverse temperature $\beta=1/T=1/160$.
Throughout the paper, the half-bandwidth $D$ of the Lorentzian
DOS of the reservoirs is fixed at $D=4$. Singular peaks, developed at 
frequencies $\omega=\Phi/2,-\Phi/2$, correspond to the Fermi 
levels of the left and right reservoirs, respectively.
As the chemical potential of the drain is lowered by $\Phi/2$,
the ionization energy of 
removing one electron from the QD ($E=\epsilon_d$)
to the Fermi level of the drain ($E=-\Phi/2$) is reduced to
$\Delta E=|\epsilon_d|-\Phi/2$. Therefore a strong
Kondo resonance forms on the electrons coming from the right-hand-side
continuum at $\omega=-\Phi/2$.
Similarly, the resonance formed from the left
reservoir produces a much suppressed resonance at $\omega=\Phi/2$.

The resulting hybridization Eq.~(\ref{eq:hyb}) due to the left and 
right reservoirs is plotted in FIG.~\ref{fig:hyb}(b). As discussed
below Eq.~(\ref{eq:hyb}), the hybridization to the source ($L$)
reservoir has strong intensity at low energy with a strong peak
at $\omega=-\Phi/2$ while it is suppressed at high energy with
depleted intensity at $\omega=+\Phi/2$. The hybridization
to the drain reservoir shows the opposite behavior.

\begin{figure}[bt]
\rotatebox{0}{\resizebox{3.0in}{!}{\includegraphics{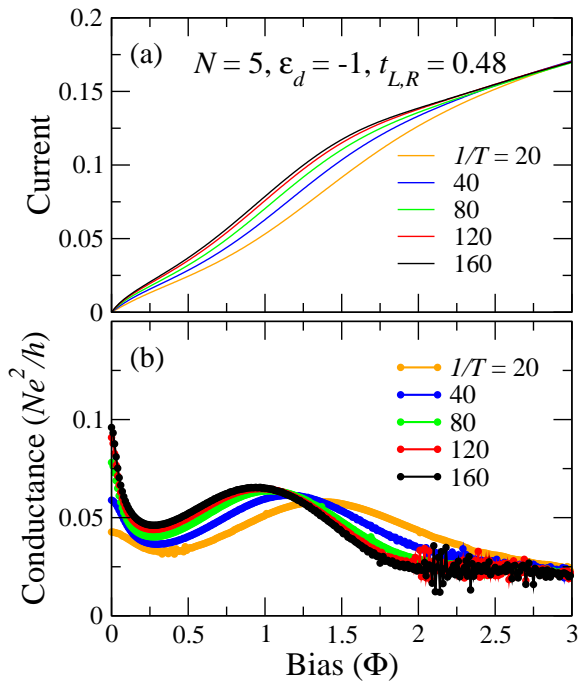}}}
\caption{
(a) $I-V$ curves (current in arbitrary unit) in the Kondo regime.
(b) Conductance $G(\Phi)=edI/d\Phi$ normalized to the unitary limit
$Ne^2/h$ (with $N$ the degeneracy in the Anderson model). 
Three peaks correspond to the Kondo resonance,
inelastic transport via the charge excitation of the QD at $\Phi
=|\epsilon_d|$, and a broad background of inelastic transport
of incoherent multiple charge excitations on the QD.
}
\label{fig:iv}\end{figure}

The $I-V$ curves for the parameter set $N=5,\ \epsilon_d=-1,\
t_{L,R}=0.48$ in the Kondo regime are shown in FIG.~\ref{fig:iv}(a)
at different inverse temperatures $1/T$. For all the tested
parameter sets, the current monotonically increases with the bias.
The differential conductance $dI/d\Phi$ in (b) shows three
distinct features: Kondo peak, inelastic current peak
at $\Phi=|\epsilon_d|$ and a very broad inelastic current at
high bias.

The Kondo peak, often referred to as the anomalous conductance peak, is
the hall-mark of strongly correlated transport. 
We emphasize that the Kondo peak arises from the combination of
the nonequilibrium Green function ${\cal G}_d(i\omega)$ and the retarded
Green function $g_d(i\omega\pm\Phi/2)$ with the strong correlation effects
present in both Green functions [see Eq.~(\ref{eq:currfinal})].
The conductance $G(\Phi)[=edI/d\Phi]$ is normalized to the unitary limit
$G_0=Ne^2/h$ with the local degeneracy $N$.

\begin{figure}[bt]
\rotatebox{0}{\resizebox{3.0in}{!}{\includegraphics{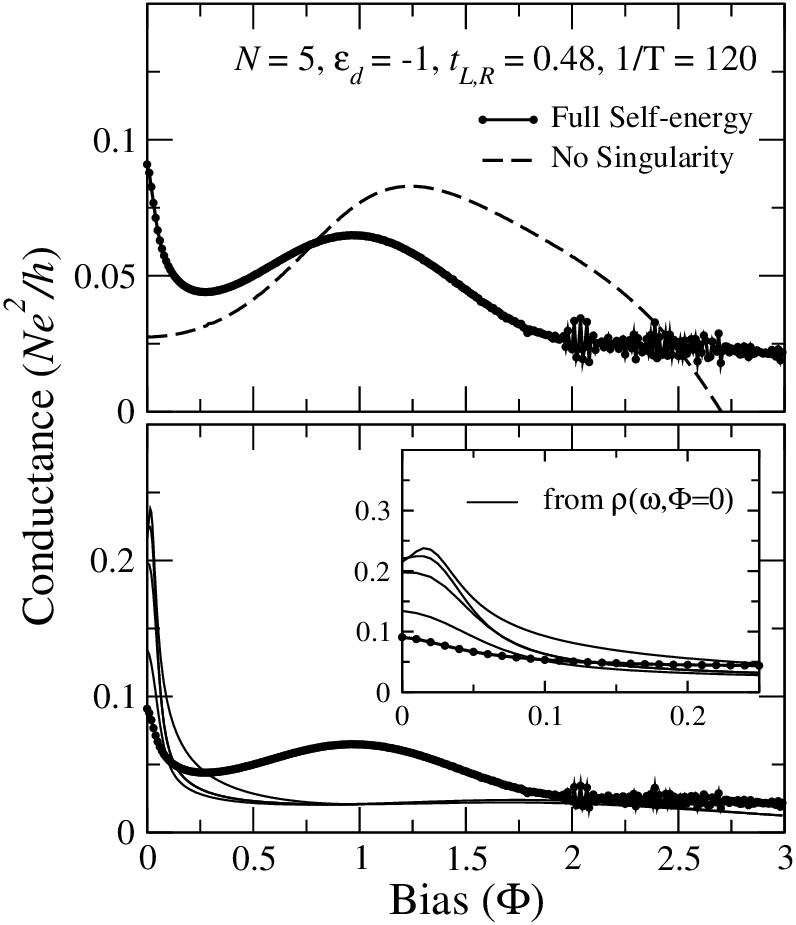}}}
\caption{(a) Normalized conductance to the unitary limit with 
(solid circles) and without (dashed line)
the strong correlation effects in the bias operator $\hat{Y}$.
With the correlation effect turned off in $\hat{Y}$, 
by replacing the Fermi-Dirac function
in Eq.~(\ref{eq:logarithm}), the conductance
does not show the Kondo peak. (b) Current calculated using
the rigid zero-bias spectral density (thin lines). See text
for details. Nonequilibrium spectral function is essential
for inelastic transport.
}
\label{fig:log}\end{figure}

To discuss the main conclusion and the limitations of this work, we 
present in FIG.~\ref{fig:log} comparisons of the results with other approaches.
First, to show that the strong correlation effects should be included 
at the level of the effective Hamiltonian $\hat{H}-\hat{Y}$, 
the logarithmic singularity 
in Eq.~(\ref{eq:logarithm}) is turned off by replacing the Fermi-Dirac 
factor $f_\alpha(\epsilon)$ by one.
The resulting $I-V$ characteristics (dashed line in (a)) does
not have the Kondo peak.
It is clear that 
even though the Kondo peak is present in the nonequilibrium Green
function ${\cal G}_d(\epsilon)$, its absence in the retarded Green
function $g_d(\epsilon)$ and therefore in $\hat{Y}$, leads to 
no zero-bias conductance peak.
This proves that the strong correlation effects in the bias operator
$\hat{Y}$ are essential to build correct steady-state nonequilibrium
ensemble. Furthermore, at large bias $\Phi\sim 2|\epsilon_f|$, not only
is the broad spectrum of inelastic transport missing but also the
differential conductance becomes negative. This underestimation
of current is due to that the
ionization spectrum of the source reservoir 
at energy $\Phi/2+\epsilon_d$ and that of the drain at
energy $-\Phi/2+\epsilon_d$ have reduced overlap.
However, proper inclusion of the retarded self-energy term compensates the
reduced overlap through a strong amplitude in the $d^0$-configuration
which mediates the inelastic tunneling.
(see the strong spectra at $\omega>0$ for $\Phi=2|\epsilon_d|$ in
FIG.~\ref{fig:rho}).

Since the Friedel-Langreth Fermi-liquid relation of the Anderson
model has the QD spectral function at zero frequency $\rho(0)=
(1/\pi\Gamma)\sin^2(\pi n_d/N)$~\cite{yamada} in the infinite-$U$
limit, the zero-bias conductance becomes
\begin{equation}
G(\Phi=0)=\frac{Ne^2}{h}\sin^2\left(\frac{\pi n_d}{N}\right).
\end{equation}
Using the typical value of $n_d=0.9$ in these calculations, we get
$G(\Phi=0)\approx 0.29(Ne^2/h)$, about 34\% larger than the
values shown in FIG.~\ref{fig:iv}.
Even though this work shows that the correlation effects
included in $\hat{Y}$ produces strongly correlated
nonequilibrium ensemble, the discrepancy of the zero-bias conductance 
is problematic. This is due to the different 
approximations used to calculate separately the retarded Green function
$g_d(\omega)$ in the scattering state 
operator Eq.~(\ref{eq:retself}), and the nonequilibrium Green
function ${\cal G}_d(\omega)$
Eqs.~(\ref{eq:selfb},\ref{eq:selff}) within the NCA. Comparing the
sharpness of the Kondo peaks in the FIGs.~\ref{fig:hyb} and \ref{fig:rho},
we can see that the correlation effects in the retarded self-energy 
Eq.~(\ref{eq:retself}) have been significantly underestimated.
In evaluating the current, Eq.~(\ref{eq:currfinal}), the peaks in the
two Green functions get convoluted in the Matsubara sum and the
over-broadened peak in $g_d(\omega)$ has the effect of smearing out
the sharp Kondo peak in ${\cal G}_d(\omega)$, leading to the underestimated
zero-bias conductance. 

It should be emphasized that this shortcoming is
due to the inconsistent approximations in the two Green functions, 
$g_d(\omega)$ and ${\cal G}_d(\omega)$, not the mapping of nonequilibrium
itself. If the same
approximation had been used for the both Green functions, for
instance by using the NCA summation scheme (see FIG.~\ref{fig:nca})
for the self-energy in the scattering state operator
(see FIG.~\ref{diagram}), the zero-bias conductance should 
have been correct.
As proved in Appendix C, the mapping of nonequilibrium
produces the correct limit of the zero-bias conductance.
Furthermore, as will be discussed in Section~\ref{sec:sc},
we develop an algorithm with one Green function and the
above problem will disappear.

\begin{figure}[bt]
\rotatebox{0}{\resizebox{3.0in}{!}{\includegraphics{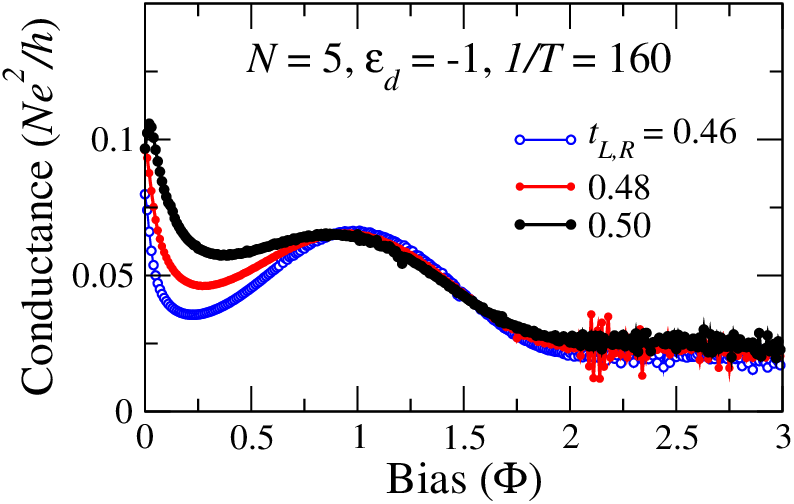}}}
\caption{Normalized conductance to the unitary limit with and without
the strong correlation effects in the bias operator $\hat{Y}$.
To turn off the correlation effect in $\hat{Y}$, the logarithmic 
singularity is suppressed by replacing the Fermi-Dirac function
in Eq.~(\ref{eq:logarithm}). The $I-V$ curve without the correlation
effects does not produce the Kondo peak in the conductance.
}
\label{fig:tx}\end{figure}

The central peak at $\Phi=|\epsilon_d|$ in FIG.~\ref{fig:iv}(b) is 
due to the inelastic transport via charge excitation. 
Analogous to the one-phonon
inelastic process in the electron-phonon coupled QDs~\cite{elph}, 
electrons tunneling from the source to the drain electrode lose
energy of the voltage drop ($\sim\Phi$) while creating charge
excitation on the QD with the ionization energy $\sim |\epsilon_d|$.

The spectral function (FIG.~\ref{fig:rho}) at $\Phi=0$ has 
a Kondo peak near the
chemical potential and the ionization energy peak at $\omega=
\epsilon_d$. If we naively expect that the current is
proportional to the integral of the zero-bias spectral function
between the source-drain chemical potentials,
the single-charge ionization process would result in an 
incorrect conductance peak at 
$\Phi=2|\epsilon_d|$, since we model that the source/drain Fermi 
levels are displaced by $\pm\Phi/2$ with respect to the QD levels.
The current calculated by assuming a rigid spectral function
$I_{\rm rigid}\propto \int d\omega\rho_d(\omega,\Phi=0)
[f(\omega-\Phi/2)-f(\omega+\Phi/2)]$, plotted as thin lines
in FIG.~\ref{fig:log}(b), does not show the inelastic peak
at $\Phi=|\epsilon_d|$. 

The broad background of inelastic transport is due to the
strong amplitude of $d^0$ configuration as noted above.
At such high bias, the system is out of the 
Kondo regime, as can be inferred from the strong
spectral weight in the positive frequency ($d^1\to d^0$)
at $\Phi=2|\epsilon_d|$ in FIG.~\ref{fig:rho}.
Despite the noise in the conductance due to the numerical 
differentiation at high bias, the trend of increasing
intensity of the inelastic transport with the tunneling
parameter (see FIG.~\ref{fig:tx})
agrees with the above observation.
We note that the infinite-$U$ approximation becomes 
increasingly unreliable at high bias
since the strong electron-channel of the 
current-influx to the QD ($d^1\to d^2$) is precluded.

Finally, FIG.~\ref{fig:tx} presents differential conductance as
the tunneling parameter is changed. As the tunneling
increases, the Kondo resonance becomes stronger and
results in pronounced zero-bias anomaly peaks. Since the
Kondo peak is shifted to the positive frequency from the 
Fermi energy by the Kondo temperature, the
conductance peak moves from zero to finite bias.

\section{Self-consistent algorithm}
\label{sec:sc}

It became clear from the
above analysis that the strong correlation effects should 
be present in the boundary condition. The main problem
in the above approach
arose from the fact that the nonequilibrium ensemble
and the time-evolution are governed by two different
operators, $\hat{H}-\hat{Y}$ and $\hat{H}$.
Therefore, to generalize the above method, it is essential
to unify the two Hamiltonians with a consistent
approximation to solve the generalized impurity model.
In this section we propose a practical algorithm to
solve this problem.

We start from the observation that the scattering state
operators, being eigen-operators of Hamiltonian, also serve
as the basis for building the bias operator. 
The Lippmann-Schwinger equation Eq.~(\ref{eq:lippmann}) is 
equivalent to the equation of motion,
\begin{equation}
(\epsilon_{\alpha k}-{\cal L}+i\eta)\psi^\dagger_{\alpha k\sigma}
=i\eta c^\dagger_{\alpha k\sigma}.
\end{equation}
From the anti-commutation relation Eq.~(\ref{eq:anticommute}),
we have
\begin{equation}
{\cal L}_Y\psi^\dagger_{\alpha k\sigma}=
[\hat{Y},\psi^\dagger_{\alpha k\sigma}]=\frac{\alpha\Phi}{2}
\psi^\dagger_{\alpha k\sigma}.
\end{equation}
Combining the above two equations, we have a new equation of
motion
\begin{equation}
(\epsilon_{\alpha k}-\frac{\alpha\Phi}{2}-{\cal L}_{H-Y}+i\eta)
\psi^\dagger_{\alpha k\sigma}=i\eta c^\dagger_{\alpha k\sigma},
\end{equation}
which is equivalent to the scattering state equation
\begin{equation}
\psi^\dagger_{\alpha k\sigma}=c^\dagger_{\alpha k\sigma}
+\frac{1}{\epsilon_{\alpha k}-\alpha\Phi/2-{\cal L}_{H-Y}+i\eta}
{\cal L}'_V c^\dagger_{\alpha k\sigma}.
\end{equation}
We now have only one Green function, propagating with the
effective Hamiltonian $\hat{H}-\hat{Y}$.
The complication is the new tunneling interaction in
${\cal L}'_V$ which is modified by the correlation terms
in $\hat{Y}$. However this effect becomes the second order
effect since the energy denominator in the Green function
has the main strong correlation effects in the modified
hybridization function in $\hat{Y}$, Eq.~(\ref{eq:hyb})
as considered in the
previous section. Furthermore, in the small bias limit
where the strong correlation effects are strongest, the
correction in ${\cal L}'_V$ is proportional to the bias
$\Phi$ and ${\cal L}'_V\approx{\cal L}_V$ becomes a
good approximation.

Therefore, we propose the following self-consistent algorithm.
\begin{enumerate}
\item Start from an approximate $\hat{Y}$, for
example from the non-interacting resonant level model~\cite{elph}.
\item Solve $\hat{H}-\hat{Y}$ for the retarded QD Green function
$G^{\rm ret}_d(\omega)$. 
\item 
Approximate ${\cal L}'_V={\cal L}_V$
and construct the new bias
operator $\psi^\dagger$ using the quasi-particle approximation,
Eq.~(\ref{eq:psitotal}), and consequently $\hat{Y}$ with the 
modified hybridization
function given by $G^{\rm ret}_d(\omega)$ [see Eq.~(\ref{eq:hyb})].
\item Go back to the step (i) until it converges.
\end{enumerate}
In this scheme we have a single Green function and
the problems arising from two Green functions, such as the 
underestimated zero-bias conductance, will not occur. 

\section{Conclusion}

We have developed a procedure of mapping a strongly correlated
steady-state nonequilibrium to an effective equilibrium.
The anti-commutation and completeness relations of the scattering state operators
are derived in the interacting limit and the bias operator
has been rewritten in a physically appealing form. 
The mapping of nonequilibrium in the zero-bias limit has
been shown to produce the correct linear-response theory. 

In the example of the infinite-$U$ Anderson impurity Hamiltonian,
as a model for Kondo dot systems, we have derived and analyzed
the scattering state operators as the basis for the statistical
operator which accounts for the nonequilibrium boundary
condition. The strong correlation effects should be included
at the Hamiltonian level in the statistical bias operator
$\hat{Y}$ to produce the zero-bias Kondo anomaly. The 
current-voltage relation has been calculated using the
equilibrium diagrammatic technique.

Despite the growing understanding of the nonequilibrium
ensemble, there are issues to be resolved for this
method to be readily applicable to general Hamiltonians.
To this end, we have developed a self-consistent algorithm
with one kind of Green functions
by exploiting the property of the scattering state
operators as eigen-operators simultaneously to the Hamiltonian
and the bias operators.

The formulation of nonequilibrium steady-state through a mapping
to an effective equilibrium not only provides an alternative way
of understanding the nonequilibrium, but also can solve various 
boundary condition problems which are otherwise hard to formulate.
For instance, thermal boundary condition, such as thermo-electric
effects under a finite temperature difference between reservoirs,
can be described quite naturally 
in the scattering state basis. 

One other interesting area is the electron transport with interactions
with classical variables which are then
self-consistently controlled by the transport, in the sense of the
Felicov-Kimball model~\cite{felicov}. Electron transport through a chain
immersed in a dielectric medium
or conduction through fluctuating (classical) spin distributions 
would be such examples.
Since non-interacting scattering states
can be written down exactly and the {\it Boltzmann factors} are
now given in a definite expression $\exp[-\beta(\hat{H}-\hat{Y})]$,
we can study transport physics in such regime with the current level
of formulation.

\begin{acknowledgments}
I thank helpful discussions with Natan Andrei, Pankaj Mehta, 
Eran Lebanon, Michael Fuda and Lingyin Zhu.
I acknowledge support from the National Science 
Foundation DMR-0426826.
\end{acknowledgments}

\appendix

\section{Proof of $\{\psi^\dagger_r,\psi_s\}=\delta_{rs}$.}

In the following discussion, we will consider the Hamiltonian
with the interaction present only in the QD, such that 
${\cal L}_Vc^\dagger_s=(t/\sqrt\Omega)d^\dagger$ where ${\cal L}_V$
includes the tunneling and many-body interaction terms. 
For simplicity, we use the subscript $s$ to denote collectively
the reservoir and continuum indices $(\alpha,k)$ and
suppress the spin index in the following. 
The Lippmann-Schwinger
Eq.~(\ref{eq:lippmann}) becomes
\begin{eqnarray}
\psi^\dagger_r & = & c^\dagger_r+\frac{t_r}{\sqrt\Omega}\frac{1}{\epsilon_r
-{\cal L}+i\eta}d^\dagger \\
\psi_s & = & c_s+\frac{t_s}{\sqrt\Omega}\frac{1}{\epsilon_s
+{\cal L}-i\eta}d, 
\end{eqnarray}
where we have used the relation $({\cal L}A)^\dagger=(HA-AH)^\dagger
=A^\dagger H-HA^\dagger=-{\cal L}A^\dagger$. We write
\begin{equation}
\{\psi^\dagger_r,\psi_s\} = \delta_{rs}+\hat{\Delta}_{rs},
\end{equation}
with
\begin{eqnarray}
\hat{\Delta}_{rs}  & = &
\left\{\frac{t_r/\sqrt\Omega}{\epsilon_r-{\cal L}+i\eta}d^\dagger,
c_s\right\} +
\left\{c^\dagger_r,\frac{t_s/\sqrt\Omega}{\epsilon_s+{\cal L}-i\eta}d
\right\} \nonumber \\
& & +
\frac{t_rt_s}{\Omega}\left\{\frac{1}{\epsilon_r-{\cal L}+i\eta}d^\dagger,
\frac{1}{\epsilon_s+{\cal L}-i\eta}d\right\}.
%
%
%
%
\label{app:delta}
\end{eqnarray}
%
%
%
%
From the relations such as
\begin{eqnarray}
& & (\epsilon_r-\epsilon_s-{\cal L}+i\eta)\left[\left(
\frac{1}{\epsilon_r-{\cal L}+i\eta}d^\dagger\right)c_s\right] \nonumber \\
& & =d^\dagger c_s +\frac{t_s}{\sqrt\Omega}
\left(\frac{1}{\epsilon_r-{\cal L}+i\eta}d^\dagger\right)d,
\label{app:a9}
\end{eqnarray}
%
%
%
%
$\hat{\Delta}_{rs}$ can be written as
\begin{eqnarray}
\hat{\Delta}_{rs} & = & \frac{t_rt_s}{\Omega}\left[
\frac{1}{\epsilon_r-\epsilon_s-{\cal L}+i\eta}
\left\{\frac{1}{\epsilon_r-{\cal L}+i\eta}d^\dagger,d\right\}\right.
\nonumber \\
& & -\frac{1}{\epsilon_r-\epsilon_s-{\cal L}+i\eta}
\left\{d^\dagger,\frac{1}{\epsilon_s+{\cal L}-i\eta}d\right\}
\nonumber \\
& & \left.+\left\{\frac{1}{\epsilon_r-{\cal L}+i\eta}d^\dagger,
\frac{1}{\epsilon_s+{\cal L}-i\eta}d\right\}
\right].
\label{app:drs}
\end{eqnarray}
Here we have used the relations $\{d^\dagger,c_s\}=0$ and
${\cal L}(\hat{A}\hat{B})=({\cal L}\hat{A})\hat{B}
+\hat{A}({\cal L}\hat{B})$.
%
%
%
The last term in the previous equation can be rewritten as
\begin{eqnarray}
& &
\left\{\frac{1}{\epsilon_r-{\cal L}+i\eta}d^\dagger,
\frac{1}{\epsilon_s+{\cal L}-i\eta}d\right\} \\
& & =\frac{-1}{\epsilon_r-\epsilon_s-{\cal L}+2i\eta} \nonumber \\
& &\times \left[
\left\{\frac{1}{\epsilon_r-{\cal L}+i\eta}d^\dagger,d\right\}
-\left\{d^\dagger, \frac{1}{\epsilon_s+{\cal L}-i\eta}d\right\}
\right],\nonumber
\end{eqnarray}
which can be easily verified by multiplying 
$(\epsilon_r-\epsilon_s-{\cal L}+2i\eta)$ to the both sides
of the above equation.
The first two terms in Eq.~(\ref{app:drs}) are canceled 
by the third term in the $\eta\to 0$ limit,
$\Delta_{rs}=0$.\cite{subtlety}
%
%
Finally, we have the desired relation
\begin{equation}
\{\psi^\dagger_r,\psi_s\}=\delta_{rs}.
\end{equation}
which corresponds to the isometry of the M{\o}ller
operator in the $S$-matrix formalism~\cite{bohm}.

\section{Boundary condition $\hat{Y}$ in terms of current
operator}

Following similar steps considered in deriving the commutation relation, we
can reduce the bias operator $\hat{Y}$ into a convenient form in terms of
current operators. We write $\hat{Y}$ as
\begin{equation}
\hat{Y}=\frac{\Phi}{2}(\hat{\cal N}_L-{\cal N}_R),\mbox{ with\ }
\hat{\cal N}_\alpha=\sum_k\psi^\dagger_{\alpha k}\psi_{\alpha k}.
\end{equation}
Similar steps leading to Eq.~(\ref{app:delta}) give
\begin{eqnarray}
& & \psi^\dagger_{\alpha k}\psi_{\alpha k} = c^\dagger_{\alpha k}c_{\alpha k}
\nonumber \\
& & +\frac{t_{\alpha k}}{\sqrt\Omega}\left[\left(\frac{1}{\epsilon_{\alpha k}-{\cal L}+i\eta}d^\dagger\right)
c_{\alpha k}+
c^\dagger_{\alpha k}\left(\frac{1}{\epsilon_{\alpha k}+{\cal L}-i\eta}d\right)
\right] \nonumber \\
& & +
\frac{t_{\alpha k}^2}{\Omega}\left(\frac{1}{\epsilon_{\alpha k}-{\cal L}+i\eta}d^\dagger\right)
\left(\frac{1}{\epsilon_{\alpha k}+{\cal L}-i\eta}d\right).
\label{app:b2}
\end{eqnarray}
The second and third terms can be rewritten as Eq.~(\ref{app:a9})
with $\epsilon_r=\epsilon_s=\epsilon_{\alpha k}$:
\begin{eqnarray}
& & \left(\frac{1}{\epsilon_{\alpha k}-{\cal L}+i\eta}d^\dagger\right)c_{\alpha k} \nonumber \\
&  & =\frac{1}{-{\cal L}+i\eta}
\left(d^\dagger c_{\alpha k}+\frac{t_{\alpha k}
/\sqrt\Omega}{\epsilon_{\alpha k}-{\cal L}+i\eta}d^\dagger\cdot
d\right).
\label{app:b3}
\end{eqnarray}
The second term in Eq.~(\ref{app:b3})
is canceled by the last term of the Eq.~(\ref{app:b2})
in the $\eta\to 0$ limit,
\begin{eqnarray}
& &
\left(\frac{1}{\epsilon_{\alpha k}-{\cal L}+i\eta}d^\dagger\right)
\left(\frac{1}{\epsilon_{\alpha k}+{\cal L}-i\eta}d\right) \\
& & =\frac{1}{{\cal L}-2i\eta} \left[
\left(\frac{1}{\epsilon_{\alpha k}-{\cal L}+i\eta}d^\dagger\right)d
-d^\dagger \left(\frac{1}{\epsilon_{\alpha k}+{\cal L}-i\eta}d\right)
\right].\nonumber
\end{eqnarray}
Following the same steps as in the previous Appendix, we have
\begin{equation}
\psi^\dagger_{\alpha k}\psi_{\alpha k} = c^\dagger_{\alpha k}c_{\alpha k}
+\frac{t_{\alpha k}}{\sqrt\Omega}\frac{1}{-{\cal L}+i\eta}
(d^\dagger c_{\alpha k}
-c^\dagger_{\alpha k}d).
\label{app:psi2}
\end{equation}
Therefore the bias operator can be expressed in a simple form as
\begin{equation}
\hat{Y} = \Phi\left[\hat{y}_0 
+\frac{1/2}{-{\cal L}+i\eta}(d^\dagger \tilde{c}_L
-\tilde{c}^\dagger_L d-d^\dagger \tilde{c}_R+\tilde{c}^\dagger_R d)\right],
\end{equation}
with $\hat{y}_0=\frac12\sum_k(c^\dagger_{Lk}c_{Lk}-c^\dagger_{Rk}c_{Rk})$ and
$\tilde{c}^\dagger_\alpha=\Omega^{-1/2}\sum_k t_{\alpha k}c^\dagger_{\alpha k}$.
Defining the source-to-drain electric current as
\begin{eqnarray}
\hat{I} & = & \frac12(\hat{I}_L-\hat{I}_R) \nonumber \\
& = & \frac{ie}{2}(d^\dagger \tilde{c}_L-\tilde{c}^\dagger_L d
-d^\dagger \tilde{c}_R-\tilde{c}^\dagger_R d),
\label{app:idef}
\end{eqnarray}
we have the bias operator in terms of the current
operator as,
\begin{equation}
\hat{Y}\equiv\Phi\hat{y} = \Phi\left[\hat{y}_0-\frac{1}{e}\frac{i}{-{\cal L}+i\eta}\hat{I}\right].
\label{app:ywithi}
\end{equation}
Or, alternatively, by making use of 
${\cal L}_V(c^\dagger_{\alpha k}c_{\alpha k})=(t_{\alpha k}/\sqrt\Omega)
(d^\dagger c_{\alpha k}-c^\dagger_{\alpha k}d)$ for Hamiltonians
with only local interactions on the quantum dots,
\begin{equation}
\hat{Y} = \Phi\left[\hat{y}_0+\frac{1}{-{\cal L}+i\eta}{\cal L}_V\hat{y}_0\right],
\end{equation}
which is equivalent to the equation of motion for the bias
operator $\hat{Y}$, $[\hat{H},\hat{Y}]=i\eta(\hat{Y}-\hat{Y}_0)$,
in Hershfield's paper~\cite{hershfield}.

\section{Derivation of zero-bias limit conductance}

In the zero-bias limit, the present formulation via mapping of nonequilibrium
should recover the linear response formula with the conductance
given by the current-current correlation function. With the expression
derived in the previous section, one can prove this as follows.
The current is evaluated via the thermal average taken over the nonequilibrium
ensemble,
\begin{equation}
I = Z^{-1}{\rm Tr}\left(e^{-\beta(\hat{H}-\hat{Y})}\hat{I}\right),
\mbox{\ with\ }Z={\rm Tr}\,e^{-\beta(\hat{H}-\hat{Y})}.
\end{equation}
By differentiating the current by the bias $\Phi$ in
$\hat{Y}=\Phi\hat{y}$ about the zero-bias limit, we obtain the expression
of the conductance in terms of the imaginary-time correlation function,
\begin{equation}
G_0 = e\frac{\partial I}{\partial\Phi}
= e\int^\beta_0 d\tau
\langle \hat{y}(\tau)\hat{I}(0)\rangle
\mbox{\ with\ }\hat{y}(\tau)=e^{\tau{\cal L}}\hat{y},
\end{equation}
where we have used the relation $\langle\hat{I}\rangle=0$.

The $\hat{y}$ term has two contributions, as in Eq.~(\ref{app:ywithi}), from
$\hat{y}_0$ and $\hat{I}$. It can be shown that the contribution from $\hat{y}_0$
can be dropped, as follows. For the sake of argument, we express
the Hamiltonian $\hat{H}$ in terms of real numbers. Since the expectation
values such as $\langle \hat{y}_0d^\dagger c_L\rangle$ are all real
numbers, they have no contributions in $\langle \hat{y}_0 \hat{I}\rangle_0$
according to Eq.~(\ref{app:idef}). Therefore we have
\begin{equation}
G_0 = -\int^\beta_0 d\tau
\langle \hat{I}_1(\tau)\hat{I}(0)\rangle,
\mbox{\ with\ }
\hat{I}_1 = \frac{i}{-{\cal L}+i\eta}\hat{I}.
\end{equation}
Analytically continuing the above integral from the imaginary-time to 
the real-time integral, we have the fluctuation-dissipation theorem
\begin{equation}
G_0  =  \int^\infty_{-\infty} dt\,\chi(t)\mbox{\ with\ }
\chi(t)=-i\theta(-t)\left\langle [\hat{I}(0),\hat{I}_1(t)]\right\rangle.
\end{equation}
Fourier transform of the response function $\chi(t)$ becomes
\begin{eqnarray}
\chi(\omega) & = & -i\int^0_{-\infty} dt\,e^{\eta t+i\omega t}
\left\langle [\hat{I},e^{i{\cal L}t}\hat{I}_1]\right\rangle \nonumber \\
& = & -i\left\langle\left[\hat{I},\frac{i}{\omega-{\cal L}+i\eta}
\frac{i}{-{\cal L}+i\eta}\hat{I}\right]\right\rangle.
\end{eqnarray}
From
\begin{equation}
\frac{i}{\omega-{\cal L}+i\eta}\frac{i}{-{\cal L}+i\eta}
=\frac{i}{\omega}\left[\frac{i}{-{\cal L}+i\eta}
-\frac{i}{\omega-{\cal L}+i\eta}\right]
\end{equation}
and
\begin{equation}
{\rm Re}
\left\langle\hat{I}\left(\frac{i}{-{\cal L}+i\eta}\hat{I}\right)\right\rangle
={\rm Re}
\left\langle\left(\frac{i}{-{\cal L}+i\eta}\hat{I}\right)\hat{I}\right\rangle,
\end{equation}
the response function is written by the current-current response function
$\chi_{II}(t)=-i\theta(t)\langle [\hat{I}(t),\hat{I}(0)]\rangle$ as
\begin{eqnarray}
\chi(\omega) & = & \frac{1}{i\omega}\chi_{II}(\omega) 
= \frac{1}{i\omega}\left\langle\left[\hat{I},\frac{1}{\omega-{\cal L}+i\eta}\hat{I}
\right]\right\rangle \nonumber \\
& = & \frac{1}{i\omega} \int^0_{-\infty} dt\,e^{\eta t+i\omega t}
(-i)\left\langle [\hat{I}(0),\hat{I}(t)]\right\rangle.
\end{eqnarray}
Finally we obtain the zero-bias conductance in terms of the linear response theory 
in agreement with the Kubo formula,
\begin{equation}
G_0 = \lim_{\omega\to 0}\chi(\omega)
=\lim_{\omega\to 0}\frac{\chi_{II}(\omega)}{i\omega}.
\end{equation}

\section{Completeness of $\psi^\dagger_{\alpha k\sigma}$}

For non-interacting models~\cite{elph} without bound states,
the scattering state operators satisfy a completeness relation
\begin{equation}
\sum_{\alpha k}\psi^\dagger_{\alpha k\sigma}\psi_{\alpha k\sigma} = 
\sum_{\alpha k} c^\dagger_{\alpha k\sigma} c_{\alpha k\sigma} 
+ d^\dagger_\sigma d_\sigma,
\label{app:complete}
\end{equation}
and the Hamiltonian is written as a sum over $\psi_{\alpha k\sigma}$ as
\begin{equation}
\sum_{\alpha k\sigma}\epsilon_{\alpha k}
\psi^\dagger_{\alpha k\sigma}\psi_{\alpha k\sigma} = \hat{H}.
\label{app:completeE}
\end{equation}
For the interacting case, it is not clear at all whether such relations
hold when $\psi^\dagger_{\alpha k\sigma}$ is made not only of 
one-particle creation operators
but also of many-particle excitations.

First we assume that the Hamiltonian has many-body interaction $\hat{H}'$
term only on the QD ($d_\sigma$) site and it takes the form
\begin{eqnarray}
\hat{H} & = & \sum_{\alpha k\sigma}\epsilon_{\alpha k}
c^\dagger_{\alpha k} c_{\alpha k}+\sum_{\alpha k\sigma}
\frac{t_{\alpha k}}{\sqrt\Omega}\left(
d^\dagger_\sigma c_{\alpha k}+c^\dagger_{\alpha k}d_\sigma\right)\nonumber \\
& + & \epsilon_d\sum_\sigma d^\dagger_\sigma d_\sigma + 
\hat{H}'(d^\dagger_\uparrow,d_\uparrow,d^\dagger_\downarrow,d_\downarrow).
\label{app:hamil}
\end{eqnarray}
We derive corresponding relations for the interacting case starting
from Eq.~(\ref{app:psi2}). After summing over the continuum variables
on both sides of the equation,
\begin{equation}
\sum_{\alpha k}\psi^\dagger_{\alpha k}\psi_{\alpha k} 
= \sum_{\alpha k} c^\dagger_{\alpha k} c_{\alpha k}
+\sum_{\alpha k}\frac{t_{\alpha k}/\sqrt\Omega}{-{\cal L}+i\eta}
\left(d^\dagger_\sigma c_{\alpha k}-c^\dagger_{\alpha k}d_\sigma\right).
\end{equation}
With the Hamiltonian Eq.~(\ref{app:hamil}), 
\begin{equation}
{\cal L}(d^\dagger_\sigma d_\sigma)=\sum_{\alpha k}
\frac{t_{\alpha k}}{\sqrt\Omega}\left(
c^\dagger_{\alpha k}d_\sigma-d^\dagger_\sigma c_{\alpha k}\right)
+{\cal L}'(d^\dagger_\sigma d_\sigma).
\end{equation}
Substituting this to the previous equation, we have
\begin{eqnarray}
\sum_{\alpha k}\psi^\dagger_{\alpha k}\psi_{\alpha k} 
& = & \sum_{\alpha k} c^\dagger_{\alpha k} c_{\alpha k}
+\frac{1}{-{\cal L}+i\eta}
\left(-{\cal L}n_\sigma+{\cal L}'n_\sigma \right)\nonumber \\
& = & \sum_{\alpha k} c^\dagger_{\alpha k} c_{\alpha k}
+d^\dagger_\sigma d_\sigma + \frac{1}{-{\cal L}+i\eta}
{\cal L}'n_\sigma.
\end{eqnarray}
In the last line we have used that
$-{\cal L}/(-{\cal L}+i\eta)=1-i\eta/(-{\cal L}+i\eta)\to 1$ and
assumed that there exist no isolated energy 
eigenstates~\cite{elph,gellmann}.
If there is an isolated state $|E_0\rangle$ outside the continuum
it cannot be constructed by the scattering states.
If $n_\sigma=d^\dagger d$ has a finite amplitude of 
$\psi^\dagger_0\psi_0$, with the creation operator $\psi^\dagger_0$
of the state $|E_0\rangle$,
it does not contribute in the $k$-summation of scattering
states, {\it i.e.} $-{\cal L}/(-{\cal L}+i\eta)(\psi^\dagger_0
\psi_0) =0$. In other words, the operator $i\eta/(-{\cal L}+i\eta)$ 
projects out the contribution of isolated states in $n_\sigma$.

If the interaction is written in terms of $d$-density operators
only, such as the Hubbard Coulomb interaction $\hat{H}'=Un_\uparrow
n_\downarrow$, ${\cal L}'n_\sigma=0.$ Therefore, we recover 
Eq.~(\ref{app:complete}) in the interacting case.

We proceed similarly for Eq.~(\ref{app:completeE}):
\begin{eqnarray}
& & \sum_{\alpha k}\epsilon_{\alpha k}
\psi^\dagger_{\alpha k\sigma}\psi_{\alpha k\sigma} 
\label{app:d7} \\
& =  &\sum_{\alpha k\sigma}\epsilon_{\alpha k}
c^\dagger_{\alpha k} c_{\alpha k} 
 + \sum_{\alpha k}\frac{t_{\alpha k}/\sqrt\Omega}{-{\cal L}+i\eta}
\epsilon_{\alpha k}
\left(d^\dagger_\sigma c_{\alpha k\sigma}-c^\dagger_{\alpha k\sigma}d_\sigma\right). \nonumber
\end{eqnarray}
Since ${\cal L}c^\dagger_{\alpha k}=\epsilon_{\alpha k}
c^\dagger_{\alpha k\sigma}
+(t_{\alpha k}/\sqrt\Omega)d^\dagger_\sigma$,
\begin{equation}
d^\dagger_\sigma({\cal L}c_{\alpha k\sigma})+({\cal L}
c^\dagger_{\alpha k\sigma})d_\sigma
=-\epsilon_{\alpha k}(d^\dagger_\sigma c_{\alpha k\sigma}
-c^\dagger_{\alpha k\sigma}d_\sigma),
\end{equation}
and
\begin{eqnarray}
& & 
d^\dagger_\sigma({\cal L}c_{\alpha k\sigma})+
({\cal L}c^\dagger_{\alpha k\sigma})d_\sigma \\
& = & 
{\cal L}(d^\dagger_\sigma c_{\alpha k\sigma}+
c^\dagger_{\alpha k\sigma}d_\sigma)
-({\cal L}d^\dagger_\sigma)c_{\alpha k\sigma}-c^\dagger_{\alpha k\sigma}
({\cal L}d_\sigma).\nonumber 
\end{eqnarray}
Therefore, Eq.~(\ref{app:d7}) can be summarized, so far, as
\begin{eqnarray}
& & \sum_{\alpha k\sigma}\epsilon_{\alpha k}\psi^\dagger_{\alpha k\sigma}
\psi_{\alpha k\sigma} 
\nonumber \\
& & = \sum_{\alpha k\sigma}\epsilon_{\alpha k}
c^\dagger_{\alpha k\sigma} c_{\alpha k\sigma}+\sum_{\alpha k\sigma}
\frac{t_{\alpha k}}{\sqrt\Omega}\left(
d^\dagger_\sigma c_{\alpha k\sigma}
+c^\dagger_{\alpha k\sigma}d_\sigma\right)\nonumber \\
& & + 
\frac{1}{-{\cal L}+i\eta}\sum_{\alpha k\sigma}\frac{t_{\alpha k}}{\sqrt\Omega}
\left[ ({\cal L}d^\dagger_\sigma)c_{\alpha k\sigma}
+c^\dagger_{\alpha k\sigma} ({\cal L}d_\sigma)\right].
\end{eqnarray}
From ${\cal L}d^\dagger_\sigma=\sum_{\alpha k}(t_{\alpha k}/\sqrt\Omega)
c^\dagger_{\alpha k\sigma}+\epsilon_d d^\dagger_\sigma
+{\cal L}'d^\dagger_\sigma$, the last term in the previous equation
becomes, before the spin summation is taken,
\begin{eqnarray}
& & \frac{1}{-{\cal L}+i\eta}\sum_{\alpha k}\frac{t_{\alpha k}}{\sqrt\Omega}
\left[ ({\cal L}d^\dagger_\sigma)c_{\alpha k\sigma}
+c^\dagger_{\alpha k\sigma} ({\cal L}d_\sigma)\right]\nonumber \\
& = & \frac{1}{-{\cal L}+i\eta}\left[
({\cal L}d^\dagger_\sigma)(-{\cal L}d_\sigma
-\epsilon_d d_\sigma+{\cal L}'d_\sigma) \right. \\
& & \left.+({\cal L}d^\dagger_\sigma
-\epsilon_d d^\dagger_\sigma-{\cal L}'d^\dagger_\sigma)
({\cal L}d_\sigma) \right] \nonumber \\
& =  & \epsilon_d d^\dagger_\sigma d_\sigma
+\frac{1}{-{\cal L}+i\eta}\left[
({\cal L}d^\dagger_\sigma)({\cal L}'d_\sigma)
-({\cal L}'d^\dagger_\sigma)({\cal L}d_\sigma)\right].\nonumber
\end{eqnarray}
By defining the non-interaction part as $\hat{H}_0$ we have
\begin{eqnarray}
& & \sum_{\alpha k\sigma}\epsilon_{\alpha k}\psi^\dagger_{\alpha k\sigma} 
\psi_{\alpha k\sigma} \\
& = & \hat{H}_0+\frac{1}{-{\cal L}+i\eta}\sum_\sigma\left[
({\cal L}d^\dagger_\sigma)({\cal L}'d_\sigma)
-({\cal L}'d^\dagger_\sigma)({\cal L}d_\sigma)\right].\nonumber
\end{eqnarray}
If the interaction is the Hubbard on-site Coulomb interaction
$\hat{H}'=Un_\sigma n_{\bar{\sigma}}$,
\begin{eqnarray}
& & -({\cal L}d^\dagger_\sigma)({\cal L}'d_\sigma)
+({\cal L}'d^\dagger_\sigma)({\cal L}d_\sigma) \nonumber \\
& = & U({\cal L}d^\dagger_\sigma)n_{\bar{\sigma}}d_\sigma
+ Un_{\bar{\sigma}}d^\dagger_\sigma({\cal L}d_\sigma)\nonumber \\
& = & U\left[({\cal L}d^\dagger_\sigma)n_{\bar{\sigma}}d_\sigma
+d^\dagger_\sigma\left\{{\cal L}(n_{\bar{\sigma}}d_\sigma)
-({\cal L}n_{\bar{\sigma}})d_\sigma\right\}\right]\nonumber \\
& = & U\left[{\cal L}(n_\sigma n_{\bar{\sigma}})-({\cal L}n_{\bar{\sigma}})
n_\sigma\right].
\end{eqnarray}
Summing over spins, the interaction term becomes
\begin{equation}
U\sum_\sigma \left[
({\cal L}d^\dagger_\sigma)({\cal L}'d_\sigma)
-({\cal L}'d^\dagger_\sigma)({\cal L}d_\sigma)\right]
={\cal L}(Un_\sigma n_{\bar{\sigma}})={\cal L}\hat{H}'.
\end{equation}
Finally, we obtain for the Anderson impurity model
\begin{equation}
\sum_{\alpha k\sigma}\epsilon_{\alpha k}\psi^\dagger_{\alpha k\sigma} 
\psi_{\alpha k\sigma} = \hat{H}_0+\hat{H}'=\hat{H},
\end{equation}
in the absence of bound states.


\begin{thebibliography}{*}
\bibitem{localized} In the narrow band limit or band with a sharp
edge, there are localized states
which cannot be included in $\psi_{\alpha k\sigma}$. However they do not
contribute to $\hat{Y}$.
\end{thebibliography}
\end{document}